\documentclass[a4paper,11pt]{article}
\pdfoutput=1 

\usepackage{jheppub} 

\usepackage[T1]{fontenc} 
\usepackage[utf8]{inputenc}
\usepackage{cancel}
\usepackage{ulem}
\usepackage{amsfonts}
\usepackage{amssymb}
\usepackage{graphicx}
\usepackage{amsmath}
\usepackage{multirow}
\usepackage{enumerate}
\usepackage{mathtools}
\usepackage{subfig}
\usepackage{color}
\usepackage{float}
\usepackage{here}
\usepackage{scalerel}
\usepackage{mathrsfs}
\usepackage{float,epsfig}
\usepackage{dcolumn}

\usepackage{graphicx}
\usepackage{bm}
\usepackage{amsmath,amssymb,amsthm}

\usepackage{xcolor}
\usepackage{tikz}
\usetikzlibrary{3d}

    {\large\bfseries Acknowledgement%
    \par\medskip\normalfont\normalsize}%
    {}%

\title{\boldmath Lense-Thirring Acoustic Black Holes : Shadows and Light }

\author{Anas El Balali$^{1}$ ,}
\author{and Alessio Marrani$^{2}$}

\affiliation{$^{1}$D\'{e}partement de physique, \'Equipe des Sciences de la mati\`ere et du rayonnement,
		ESMaR\\ \small   Facult\'e des Sciences, Universit\'e Mohammed V de Rabat,  Rabat,  Morocco}
\affiliation{$^{2}$Centre for Mathematics and Theoretical Physics,University of Hertfordshire, AL10 9AB Hatfield, UK}
\emailAdd{anas.elbalali@gmail.com}
\emailAdd{a.marrani@herts.ac.uk}

\abstract{We introduce the \textit{Lense-Thiring Acoustic Black Hole} (LTABH),
motivated by the relevance of analogue models for black holes embedded in
various physical systems, such as the cosmological microwave background or
quantum superfluids. We investigate the LTABH spacetime geometry, showing
that the roots of the metric function determine a partition of the spacetime
into four regions, depending on the acoustic parameter $\xi $ (whereas the
dependence vanishes for the rotation parameter $a$); on the other hand, the
parameter $a$ turns out to affect the critical radii associated to the
maxima of the effective potential. All in all, both the acoustic sphere
radius $r_{as}$ and the photon sphere radius $r_{ps}$, respectively giving
rise to the acoustic shadow $R_{as}$ and to the optical shadow $R_{s}$,
depend on $\xi $ and $a$. More precisely, the rotation parameter $a$ is more
relevantly affecting $R_{s}$ (through a right shift), while $R_{as}$ retains
its circular shape. For what concerns the acoustic parameter, we notice that
the higher $\xi $ is, the larger the size of both shadows. All of these
results are confirmed through a detailed analysis of the distortions and of
the shadows radii. Moreover, by deriving the magnitude of the precession
frequency $\Omega $, we observe that it significantly increases near the
acoustic horizons, both in the extremal and in the non-extremal cases, which
implies that the Lense-Thirring (frame dragging) effect, which can be traced
back to $\xi $ itself, becomes important near such regions. On the other
hand, we also show that there are regions of the LTABH spacetime in which $%
\Omega $ vanishes, suggesting that therein possible probe particles would
not be affected by the frame dragging at all. Finally, we derive the
deflection of the light near the LTABH.
}

\begin{document} 
\maketitle
\flushbottom

\section{Introduction}

In Einstein's theory of General Relativity, the presence of matter in the
universe dictates how the spacetime shapes, as described by Einstein's field
equations \cite{Ip11}. Among the solutions to these equations, we find
\textit{black holes}. These objects are the result of the gravitational
collapse of massive stars, at the end of their life cycle \cite{Ip12, Ip13}.
Such a phenomenon leads to the formation of a \textit{singularity},
surrounded by the event horizon, `the sphere of no return'. Black holes are
therefore characterized by a very intense gravity that influences the
surrounding spacetime, and which may thus lead to interesting optical
phenomena. Indeed, it can be shown that black holes can act as a lens,
causing the deflection of light (\textit{gravitational lensing}) \cite{Ip14,
Ip15, Ip16, Ip17, Ip18, Ip19, Ip110, Ip111}. From another perspective, the
so-called \textit{black hole shadow} is also a consequence of the immense
gravitational pull of these objects \cite{Ip112, Ip113, Ip114, Ip115, Ip116,
Ip117}.

Furthermore, black holes exhibit thermodynamic proprieties such as the
entropy, the temperature, and most importantly the Hawking radiation, in
which quantum effects are responsible for the black hole to emit radiation
and slowly evaporate \cite{Ip118, Ip119, Ip120, Ip121, Ip122, Ip123, Ip124,
Ip125}. Along this and other research venues, the study of black holes keep
providing remarkable insights on the unification of quantum mechanics and
gravity \cite{Ip126, Ip127, Ip128, Ip129, Ip130}. Black holes have also
played an important role in the understanding of gravitational waves : the
detection of $GW\,150914$ published by LIGO and Virgo Collaboration has
marked the first gravitational wave detection resulting from the merger of a
binary black hole system \cite{Ip131}. It should also be recalled that
imaging by the Event Horizon Telescope have provided an undeniable evidence
of the existence of black holes themselves \cite{Ip132, Ip133, Ip134, Ip135,
Ip136, Ip137, Ip138, Ip139}. Such discoveries and unprecedented advances in
the observation of black holes have aroused the interest of physicists
around the world leading to a relentless and ever growing research activity.

Notwithstanding such ground-breaking results in the field of black hole
physics, the experimental detection of the black hole features is still
subject to limiting technical limitations. This can be regarded as one of
the motivations which led to the birth of a new field of research in
gravity, the so-called \textit{analogue gravity}, aiming at mimicking
gravitational pheonomena in the laboratory; needless to say, analogies have
been constantly playing a key role in mathematics and physics since they
have offered new perspectives and simplifications of complex ideas. In
particular, it is here worth remarking that, through table top experiments,
analogue gravity established key connections between rotating black holes
and the Hawking radiation. This was firstly reported in the seminal paper by
Unruh \cite{Ip21}, in which hydrodynamical flows were used to mimic some proprieties of
black holes. In this framework, sound waves which are trapped
inside a horizon led to the introduction of the name \textit{acoustic black
hole} or \textit{dumb hole}. All in all, it has been argued that \textit{any}
fluid can ultimately give birth to an acoustic black hole, if it moves
faster than the local sound velocity within a spherical surface \cite{Ip22,
Ip23,If1,If2}. Black hole analogues have been investigated within a wide range of
classical and quantum systems; just to cite a few, gravity waves \cite{Ip24}%
, water \cite{Ip25}, `slow' light \cite{Ip26,Ip27,Ip28}, optical fibers \cite%
{Ip29}, and electromagnetic waveguides \cite{Ip210}. Other models exhibiting
promising features include the superfluid helium II \cite{Ip211}, atomic
Bose-Einstein condensates \cite{Ip212}, and one-dimensional Fermi-degenerate
non-interacting gases \cite{Ip213}. For further detail and a quite complete
list of references, we address the interested reader to the comprehensive
review \cite{Ip214}.

Acoustic black holes have long been the subject of many studies. In a
thermodynamical framework, Hawking-like radiation emitted by acoustic black
holes has been successfully detected and studied in \cite{Ip215, Ip216,
Ip217}, and entropy of acoustic black holes has also been investigated in a
number of studies \cite{Ip2171, Ip2172, Ip2173, Ip2174, Ip2175}. More
insights were gained by focussing on two-dimensional systems \cite{Ip2176, Ip21761},
in which the investigation of the Bose-Einstein condensates provided an
analogue of the Bekenstein-Hawking entropy \cite{Ip2177}. While in \cite%
{Ip2178} the entanglement entropy of acoustic black holes was investigated,
the relation between the area of the acoustic black hole event horizon an
the entropy has been the subject of \cite{Ip2179}. Theoretically, the
consideration of relativistic Gross-Pitaevski and Yang-Mills theories led to
the embedding of the acoustic black hole into a curved background \cite%
{Ip218}; for instance, in \cite{Ip219} an acoustic black hole solution was
related to a black $D3$-brane. Moreover, the investigation of such black
hole analogues in the Abelian Higgs model led to the discovery of a
relativistic acoustic black hole solution in Minkowski spacetime \cite%
{Ip22, Ip221}.

An interesting property of acoustic black holes goes under the name of
\textit{acoustic shadow} : it can be regarded as the acoustic analogue of
the optical shadow of a black hole that has been captured by the EHT
collaboration, and it has been studied in a variety of backgrounds and
contexts; see e.g. \cite{Ip32, Ip33}. The deflection angle has also
been scrutinised in proximity of an acoustic black hole, in \cite{Ip34, Ip35}%
. At any rate, realistic models of acoustic black holes must necessarily
include rotation : this is one of the motivations of \cite{main}, in which a
slowly rotating solution has been constructed and investigated. In this
framework, the interest of the analogue gravity approach towards the
relativistic Gross-Pitaevskii theory as well as towards Yang-Mills (and
thus, non-Abelian) gauge theories relies on a number of reasons; here, we
will list three of them. Firstly, realistic models of astrophysical black
holes include the embedding into the cosmological microwave background, and
this makes the acoustic black holes natural candidates to be investigated in
this venue \cite{Ip218}. Secondly, the relativistic and transonic accretion
disks, surrounding supermassive black holes at the center of galaxies, serve
as unique examples of analogue gravity models present in nature \cite{Ip331,
Ip332, Ip333, Ip334}. Thirdly, one shoud recall that black holes may also
be surrounded by some quantum superfluids whose condensation can generally
give rise to analogue gravitational phenomena; for instance, such fluids have
been proposed to determine also dark matter phenomenology \cite{Ip335}.

Motivated by this wealth of facts and results, it is our belief that
acoustic black holes deserve further investigation, and they ultimately have
a chance to be detected in reality. By taking advantage of a slowly rotating
solution at hand, we aim at investigating more realistic frameowrks, in which
both phonon and photons can propagate in the corresponding spacetime.\bigskip

In this work, we will present a detailed investigation of the impact of the
rotation parameter on the slowly rotating and curved acoustic black hole. In
particular, we will study the partition of the black hole geometry into
different regions, depending on the acoustic parameter itself. By
considering the null geodesics of the corresponding spacetime, we will gain
insights about the behavior of the shadows and the corresponding distortion.
Furthermore, we will study the Lense-Thirring effect, responsible for the
frame dragging of rotating black holes. \textit{Last but not least}, we will
provide a careful analysis of the deflection angle, by exploiting the
Gauss-bonnet method \footnote{%
	In this work, we use dimensionless units, i.e. $c=G=1$.}.\bigskip

The plan of the paper is as follows. Sec. %
\ref{II} reviews of the slowly rotating acoustic black hole solution, whose
geometry is analyzed in Sec. \ref{III}. Sec. \ref{iv} is then devoted to the
study of the shadow and of the null geodesics of the black hole. We will
define an \textit{effective potential} as a function of the relevant black
hole parameters, and we will study its critical points in an analytic and
graphical way, focussing on the emergence of various critical radii. In this
framework, we will investigate the shadows as parametric functions, and
discuss their features. The subsequent Secs. \ref{v} and \ref{vi} will be
devoted to the analysis of the frame dragging effect, and to deflection of
light around an acoustic black hole, respectively. Finally, in Sec. \ref{VII}
we will summarize our results, provide remarks and an outlook to further
developments. Two appendices conclude the paper : in App. \ref{app1} the
placement of the aforementioned critical radii is discussed, whereas in App. %
\ref{app 2} the explicit, lengthy expression of a matrix relevant for the
LTABH metric is reported.

\section{Notation and Conventions}

In this paper, we adopt the following notations
\begin{table}[h!]
\caption{{\protect\footnotesize List of symbols in the paper}}
\label{nota}\centering
\begin{tabular}{|c|l|}
\hline
Symbol & Description \\ \hline
$c_s$ & The speed of sound in the acoustic medium \\ \hline
$M$ & The mass parameter of the non-acoustic case \\ \hline
$a$ & The rotation parameter \\ \hline
$\xi$ & The acoustic parameter \\ \hline
$r_H$ & The horizon radius \\ \hline
$r_{ac_-}$ & The inner acoustic critical radius \\ \hline
$r_{ac_+}$ & The outer acoustic critical radius \\ \hline
$r_{ac}$ & The acoustic critical radius in the extremal case $(\xi=4)$ \\ \hline
$E$ & The particle's energy \\ \hline
$L$ & The particle's angular momentum \\ \hline
$\mathcal{K}$ & The separation constant \\ \hline
$r_{ps}$ & The photon sphere radius in the acoustic case \\ \hline
$r_{as}$ & The acoustic sphere radius \\ \hline
$r_{ph}$ & The photon sphere radius in the non-acoustic case \\ \hline
$\eta$, $\zeta$ & The impact parameters \\ \hline
$R_{s}$ & The shadow radius \\ \hline
$R_{as}$ & The acoustic shadow radius \\ \hline
$\delta_{s}$ & The deformation of the apparent shadow \\ \hline
$\delta_{as}$ & The deformation of the acoustic shadow \\ \hline
$K^\alpha$ & The Killing vector \\ \hline
$\Omega$ & The Lense-Thirring precession frequency \\ \hline
$K_\mathcal{G}$ & The Gaussian optical curvature \\ \hline
\end{tabular}%
\end{table}

\section{Review of Lense-Thirring acoustic black hole solution}

\label{II}

In this part of the paper, we review the procedure to derive a slowly
rotating acoustic black hole solution. In order to establish this solution,
we follow the approach outlined in paper \cite{main}. Indeed, within the
framework of analogue gravity, the Lagrangian of relativistic
Gross-Pitaevskii (GP) theory of a complex scalar field $\varphi $ is written as
\begin{equation}
\mathcal{L}=g^{\mu \nu }\partial _{\mu }\varphi ^{\star }\partial _{\nu
}\varphi -m^{2}\varphi ^{\star }\varphi +\frac{b}{2}\left( \varphi ^{\star
}\varphi \right) ^{2},
\end{equation}%
which yields the following action
\begin{equation}
S=\int d^{4}x\sqrt{-g}\,\left( \big\vert\partial _{\mu }\varphi \big\vert%
^{2}-m^{2}\big\vert\varphi \big\vert^{2}+\frac{b}{2}\big\vert\varphi %
\big\vert^{4}\right) ,
\end{equation}%
where $b$ represents a coupling constant and $m$ is a parameter that depends
on the Hawking-Unruh temperature $T$ of the resulting acoustic solution
carrying information about the black hole and the acoustic metric. It is
worth noting that such a parameter can be expressed as $m^{2}\sim T-T_{c}$
with $T_{c}$ being the critical temperature where phase transitions can
occur. The variation of the action, can provide the motion of $\varphi $
which is described by the following equation
\begin{equation}
\square \varphi +m^{2}\varphi -b\big\vert\varphi \big\vert^{2}\varphi =0.
\label{KG}
\end{equation}%
The acoustic black hole solution of such a theory can be derived by
perturbing the complex scalar field $\varphi $ around the spacetime
background. Such a field can propagate in a fixed background spacetime
described by the metric
\begin{equation}
ds^{2}=-g_{tt}dt^{2}+g_{rr}dr^{2}+g_{\theta \theta }d\theta ^{2}+g_{\phi
\phi }d\phi ^{2}+2g_{t\phi }dtd\phi .  \label{first}
\end{equation}%
It is worth recalling that such a metric form has been chosen to align with
a slowly rotating black hole, similar to the \textbf{Lense-Thirring Black
Hole (LTBH)}, which will be introduced later \cite{main3}. Indeed, the
slowly rotating approximation assumes that higher-order terms of the
rotation parameter $a$, which is embedded in the component $g_{t\phi }$, are
negligible. The scalar field $\varphi $ can be expressed as a function of the
fluid density $\rho =\rho _{0}+\epsilon \rho _{1}$ and the phase $\vartheta
=\vartheta _{0}+\epsilon \vartheta _{1}$ as follows
\begin{equation}
\varphi =\sqrt{\rho \left( \vec{x},t\right) }\exp \left( i\vartheta \left(
\vec{x},t\right) \right) .  \label{EQQ1}
\end{equation}%
In the hydrodynamic or Thomas-Fermi limit the quantum pressure $P_{%
\mathcal{Q}}=\frac{\square \sqrt{\rho }}{\sqrt{\rho }}$ can be neglected in
front of the quantities $m^{2},b\rho _{0},v^{\mu }v_{\mu }$. Thus,
replacing $\varphi $ in equation \eqref{KG}, one finds the background fluid
density leading order equation which reads
\begin{equation}
b\rho _{0}\equiv 2c_{s}^{2}=m^{2}-g^{\mu \nu }v_{\mu }v_{\nu },
\label{velocity}
\end{equation}%
where $\left( \rho _{0},\vartheta _{0}\right) $ and $\left( \rho
_{1},\vartheta _{1}\right) $ are associated to the background solution in
fixed spacetime and to fluctuations respectively, while $c_{s}$ is the speed
of sound. It is worth noting that the background four-velocity $v_{\mu }$
has been defined as $v_{\mu }=(-\partial _{t}\vartheta _{0},\partial
_{i}\vartheta _{0})$. In order to determine the relativistic wave equation,
we use the continuity equation that reads as
\begin{equation}
\nabla _{\mu }\left( \rho v^{\mu }\right) =\nabla _{\mu }\left( \rho
_{0}v_{0}^{\mu }\right) +\epsilon \nabla _{\mu }\left( \rho _{0}\partial
^{\mu }\vartheta _{1}+\rho _{1}v_{0}^{\mu }\right) =0,
\end{equation}%
where $\nabla _{\mu }\left( \rho _{0}v_{0}^{\mu }\right) =0$ is associated
to the background continuity equation and the first order term gives the
linearized continuity equation for fluctuations
\begin{equation}
\nabla _{\mu }\left( \rho _{0}\partial ^{\mu }\vartheta _{1}+\rho
_{1}v_{0}^{\mu }\right) =0.  \label{eqq4}
\end{equation}%
At subleading order in $\epsilon $ \cite{main2}, the decomposition of the
scalar field $\varphi $ when replaced in equation of motion of the scalar
field \eqref{KG} leads to
\begin{equation}
\rho _{1}=-\frac{2}{b}v_{0}^{\mu }\partial _{\mu }\vartheta _{1}.
\end{equation}%
Replacing in equation \eqref{eqq4}, one gets
\begin{equation}
\nabla _{\mu }\left( \rho _{0}g^{\mu \nu }-\frac{2}{b}v_{0}^{\mu }v_{0}^{\nu
}\right) \partial _{\nu }\vartheta _{1}=0.
\end{equation}%
As a result, the effective metric can be taken as
\begin{equation}
\mathcal{G}^{\mu \nu }\propto \rho _{0}g^{\mu \nu }-\frac{2}{b}v_{0}^{\mu
}v_{0}^{\nu }.
\end{equation}%
Finally, the relativistic wave equation that dictates the propagation of the
phase fluctuations is
\begin{equation}
\frac{1}{\sqrt{-\mathcal{G}}}\partial _{\mu }\left( \sqrt{-\mathcal{G}}\,%
\mathcal{G}^{\mu \nu }\partial _{\nu }\vartheta _{1}\right) =0.
\end{equation}%
As a results, one could extract from such equation the effective metric $%
\mathcal{G}_{\mu \nu }$ which is given in App. \ref{app 2}.

Assuming that $v_{t}\neq 0$, $v_{r}\neq 0$, $v_{\theta }=0$, $v_{\phi }=0$
and $g_{rr}g_{tt}=-1$, and using the following coordinate transformaions
\begin{align}
dt& \rightarrow dt+\frac{v_{r}v_{t}}{g_{tt}\left(
c_{s}^{2}-v_{i}v^{i}\right) }dr, \\
d\phi & \rightarrow d\phi -\frac{g_{t\phi }\left(
c_{s}^{2}-v_{t}v^{t}\right) v_{t}v_{r}}{g_{\phi \phi }g_{tt}\left(
c_{s}^{2}-v_{\mu }v^{\mu }\right) \left( c_{s}^{2}-v_{r}v^{r}\right) }dr,
\end{align}%
we get the line element of a slowly rotating and curved acoustic black hole
\begin{equation}
ds^{2}=c_{s}\sqrt{c_{s}^{2}-v_{\mu }v^{\mu }}\left( \frac{%
c_{s}^{2}-v_{r}v^{r}}{c_{s}^{2}-v_{\mu }v^{\mu }}g_{tt}dt^{2}+\frac{c_{s}^{2}%
}{c_{s}^{2}-v_{r}v^{r}}g_{rr}dr^{2}+g_{\theta \theta }d\theta ^{2}+g_{\phi
\phi }d\phi ^{2}+2\frac{c_{s}^{2}-v_{t}v^{t}}{c_{s}^{2}-v_{\mu }v^{\mu }}%
g_{t\phi }dtd\phi \right) .  \label{fluidmetr}
\end{equation}%
In order to fully define such a spacetime, we identify the fluid
four-velocity with the LTBH whose metric is written as
\begin{equation}
ds^{2}=-f(r)dt^{2}+\frac{dr^{2}}{f(r)}+r^{2}d\theta ^{2}+r^{2}\sin
^{2}\theta d\phi ^{2}-2\left( 1-f(r)\right) a\sin ^{2}\theta dtd\phi ,
\label{Lens}
\end{equation}%
where the metric function is
\begin{equation}
f(r)=1-\frac{2M}{r},  \label{RNf}
\end{equation}%
and $a$ is the rotation parameter \cite{main3}. Regarding the four-velocity,
one can use the equation \eqref{velocity} and rescale in terms of natural
units of $2c_{s}^{2}$ as follows
\begin{equation}
\tilde{v}_{\mu }\rightarrow \frac{v_{\mu }}{\sqrt{2c_{s}^{2}}},
\end{equation}%
which gives
\begin{equation}
\tilde{v}_{\mu }\tilde{v}^{\mu }=\frac{m^{2}}{2c_{s}^{2}}-1.
\end{equation}%
Thus, for $m^{2}\rightarrow 0$ in the limit of the critical temperature $%
T_{c}$, we get
\begin{equation}
v_{\mu }v^{\mu }=-1.
\end{equation}%
The escape velocity of an observer who remains stationary at the radial
point $r$ can be associated with the radial component of the fluid
four-velocity. In this way, we have
\begin{equation}
v_{r}\sim \sqrt{\left( 1-f(r)\right) \xi },  \label{vr}
\end{equation}%
where $\xi $ represents an acoustic parameter that constrains the acoustic
event horizon to be outside the event horizon of the black hole. On another
hand, we can write
\begin{equation}
v_{\mu }v^{\mu }=g^{tt}v_{t}^{2}+g^{rr}v_{r}^{2}=-\frac{v_{t}^{2}}{f(r)}%
+f(r)v_{r}^{2}.  \label{eqq5}
\end{equation}%
With $v_{\mu }v^{\mu }=-1$, equation \eqref{eqq5} gives
\begin{equation}
v_{t}=\sqrt{f(r)+\left( 1-f(r)\right) f(r)^{2}\xi }.
\end{equation}%
Finally, the metric of equation \eqref{fluidmetr} can be rewritten as
\begin{equation}
ds^{2}=-F(r)dt^{2}+\frac{dr^{2}}{F(r)}+r^{2}d\theta ^{2}+r^{2}\sin
^{2}\theta d\phi ^{2}-2\left( 1-F(r)\right) a\sin ^{2}dtd\phi ,  \label{last}
\end{equation}%
where the acoustic metric functions $F(r)$ is given by
\begin{equation}
F(r)=f(r)\left[ 1-f(r)\left( 1-f(r)\right) \xi \right] .  \label{F}
\end{equation}%
For simplicity reasons, the speed of sound has been fixed, i.e $c_{s}^{2}=1/%
\sqrt{3}$. It is worth noticing that, the escape velocity $v_{r}$ in
equation \eqref{vr} implies that the parameter $\xi $ must be strictly
positive. Besides, when the acoustic parameter $\xi \rightarrow 0$, we recover
the LTBH while when $\xi \rightarrow \infty $, i.e $v_{r}\rightarrow \infty $%
, the entire spacetime is covered by the LTABH.

The spacetime geometry of a LTABH is described by equations \eqref{last}, %
\eqref{F}, and $f(r)=1-2M/r$. Such scenario offers valuable insight which
deserves further investigation. In what follows, we analyse the geometry of
the this solution, discuss the roots of the acoustic metric function $F(r)$
in terms of the acoustic parameter $\xi$ and derive the shadow equation. We
also investigate the critical radii involved in the acoustic case and shed
light on the frame dragging effect and the deflection angle of the present
solution.

\section{Geometrical analysis}

\label{III} 
\begin{figure}[h]
\begin{tabular}{ll}
\includegraphics[scale=0.32]{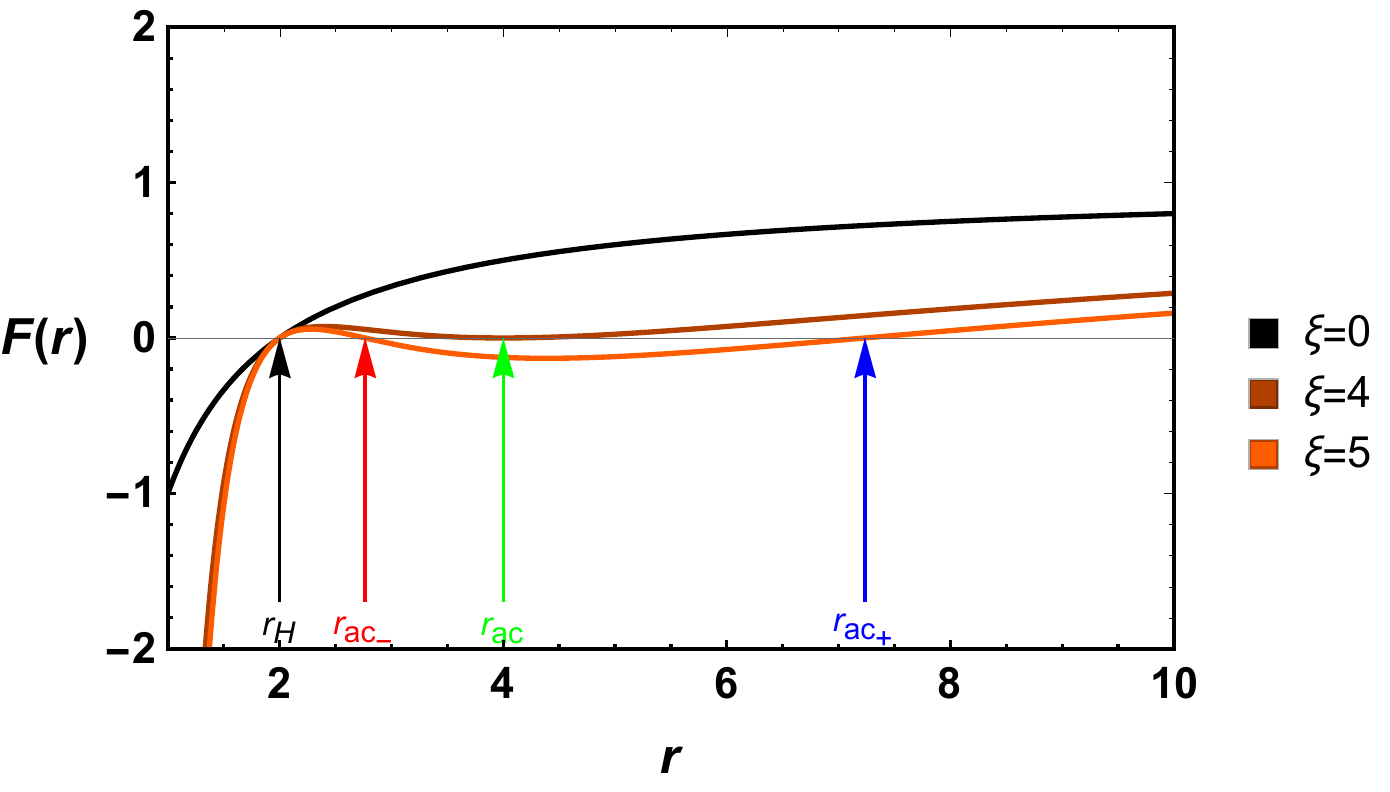} & \includegraphics[scale=0.3]{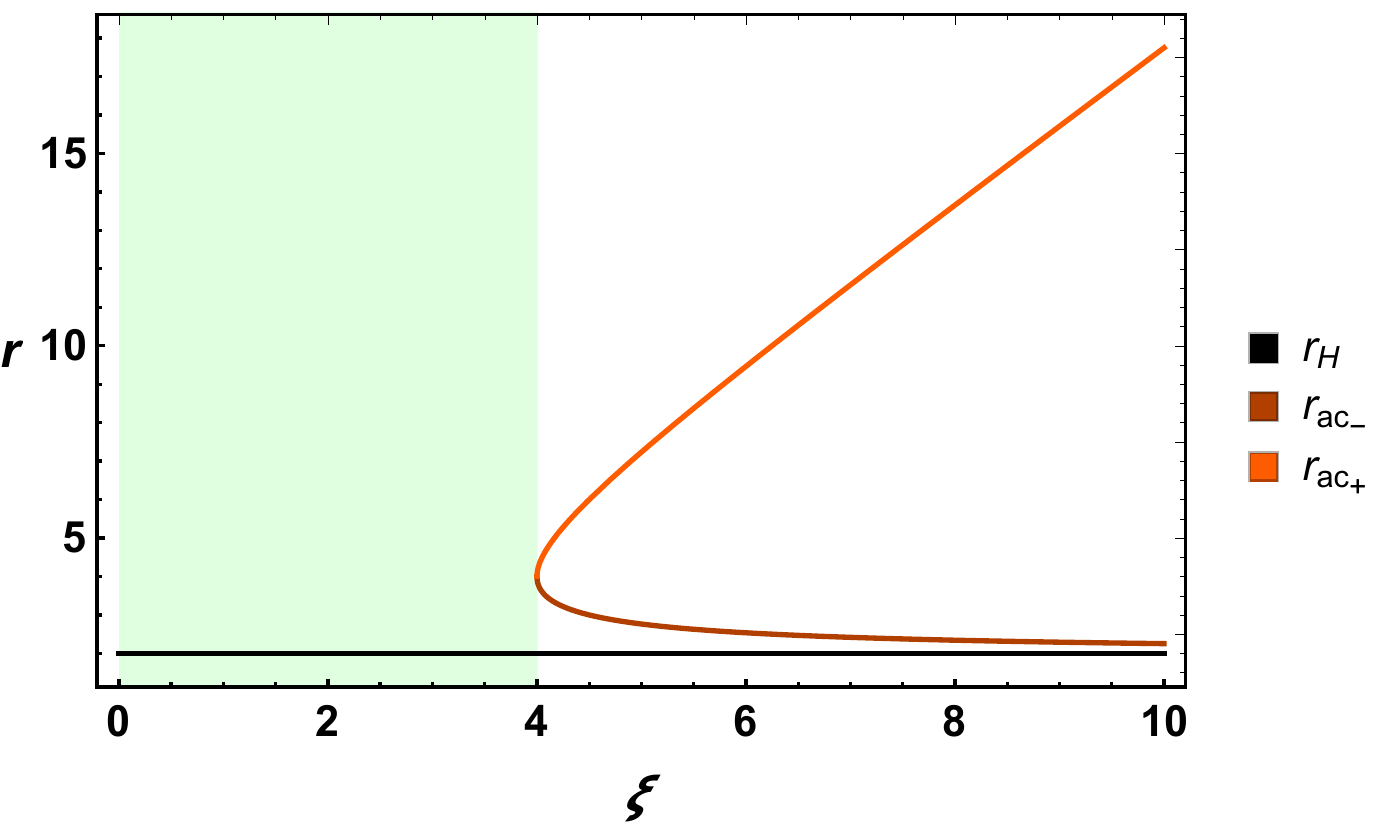}%
\end{tabular}%
\caption{ {\protect\footnotesize Left: Profile of the metric function F(r)
with its roots for different values of the parameter $\protect\xi$. Right:
The roots as a function of the parameter $\protect\xi$.
The
green region represents the non allowed values of
$\xi$, i.e $\xi < 4$. We take $M=1$. }}
\label{F-sol}
\end{figure}

The considered solution is similar to the LTBH since the event
and acoustic horizons do not depend on the rotation parameter $a$. Indeed,
by solving $F(r)=0$ three solutions are derived, which are
\begin{equation}
r_H=2M, \quad r_{ac_{-}}=M \left( \xi - \sqrt{\xi (\xi -4)} \right), \quad
r_{ac_{+}}=M\left( \xi + \sqrt{\xi (\xi -4)} \right),
\end{equation}
which are represented with the metric function $F(r)$ in figure \eqref{F-sol}%
.

By analysing the roots of the metric function $F(r)$, we remark that the
inner acoustic horizon $r_{ac_{-}}$ and the outer acoustic horizon $%
r_{ac_{+}}$ exist only if $\xi > 4 $, while
when $\xi = 4$,
we have $r_{ac_{-}}=r_{ac_{+}}=4M$ which corresponds to the extremal case of
the slowly rotating acoustic black hole. In the right side of figure %
\eqref{F-sol}, we notice that the solutions associated with the acoustic
aspect of the black hole $r_{ac_{-}}$ and $r_{ac_{+}}$ are larger than the
horizon radius $r_H$. Moreover, it is clear that the inner acoustic horizon $%
r_{ac_{-}}$ converges to the value $2M$ as the acoustic parameter $\xi$
increases, while the outer acoustic horizon $r_{ac_{+}}$ diverges to
infinity for higher values of $\xi$. This shows that as the acoustic parameter
increases, no sound wave could escape the spacetime, while other particles
can be absorbed by the event horizon. For the particular value of $\xi=4$,
we obtain the extremal case which is characterized by only one acoustic
horizon. It is worth mentioning that the positive values of $F(r)$ indicates
that the fluid velocity is less than the speed of sound which means that the
sound waves could move freely in this spacetime region. In contrast, sound
waves in a region with a negative $F (r)$ will be confined and unable to be
heard by outside observers.

Interestingly, for $\xi > 4 $ and an observer in the region $r_H < r <
r_{ac_{-}}$, the radius $r_{ac_{-}}$ could be regarded as the white hole's
acoustic horizon. In particular, the fluid flow is approaching the observer
in the subsonic area $r_H < r < r_{ac_{-}}$ from the supersonic region ( $%
r_{ac_{-}} < r < r_{ac_{+}}$), having crossed the horizon ($r_{ac_{-}}$). It
is fair to consider $r_{ac_{-}}$ to be the horizon of the acoustic white
hole for this observer because, as a result, sound waves in the subsonic
zone cannot travel against the fluid flow into the supersonic region,
whereas the opposite is permitted. Furthermore, the behavior of the metric
function $F(r)$ varies depending on $r_{ac_{-}}$ and $r_{ac_{+}}$ as it can
be noticed from the left plot of figure \eqref{F-sol}. Indeed, $F (r) > 0$
in the region of $r < r_{ac_{-}}$, while $F (r) < 0$ in the region of $r >
r_{ac_{-}}$, suggesting that the inner acoustic horizon share some
characteristics with the charged black hole's Cauchy horizon. Conversely,
the metric function near the outer acoustic horizon exhibits behavior
similar to the black hole's event horizon, with $F (r) < 0$ for $r <
r_{ac_{+}}$ and $F (r) > 0$ for $r > r_{ac_{+}}$.

Based on the aforementioned factors, we can conclude that the acoustic black
hole is characterized by four regions.

\begin{itemize}
\item Region I with $r < r_H$: inside the LTABH black hole, where neither
light nor sound waves may escape.

\item Region II, where $r_H < r < r_{ac_{-}}$, is the area where light and
sound waves can exit but are invisible to observers outside of it.

\item Region III with $r_{ac_{-}} < r < r_{ac_{+}}$, whereas the sound wave
cannot escape from but light can.

\item Region IV with $r > r_{ac_{+}}$, both the light and sound waves can
escape.
\end{itemize}

For the particular extremal case of $\xi =4$, $r_{ac_{-}} =
r_{ac_{+}}=r_{ac} $, we are left with only three regions and the function $%
F(r)$ is negative only in region I.

\section{Shadows and null geodesics}

\label{iv} In this section, we analyse the effective potential behavior of
the slowly rotating acoustic black hole described by the metric \eqref{last}%
. With the help of null geodesics, we derive relevant regions in which
the effective potential has a particular behavior. Moreover, we investigate
the shadows in the vicinity of an acoustic black hole and illustrate its
behavior.

\subsection{General approach}

Studying the null geodesics is necessary in order to inspect the evolution
of the massless particles surrounding the black hole under consideration
\cite{1S1}. In order to do so, we use the following Hamilton-Jacobi equation
\begin{equation}
\frac{\partial S}{\partial \tau}=-\frac{1}{2} g^{\mu \nu} \frac{\partial S}{%
\partial x^\mu} \frac{\partial S}{\partial x^\nu},  \label{HJ}
\end{equation}
where $\tau$ denotes the affine parameter of the null geodesic and the
Jacobi action can be separated in the following manner
\begin{equation}
S=\frac{1}{2} m_0^2 \tau-E t + L \phi+S_r\left( r \right)+S_\theta \left(
\theta \right).  \label{S}
\end{equation}
The mass $m_0$ is equal to zero for the class of massless particles. The
particles energy and angular momentum are denoted by $E$ and $L$,
respectively, and the functions $S_r\left( r \right)$ and $S_\theta \left(
\theta \right)$ only depend on $r$ and $\theta$, respectively. The following
equation is obtained using the slow rotation metric \eqref{last} by
substituting the Jacobi action \eqref{S} into the Hamilton-Jacobi equation %
\eqref{HJ}
\begin{align}
0= & -F(r)^{-1} \left( \frac{\partial S}{\partial t} \right)^2+\frac{1}{r^2}
\left( \frac{\partial S}{\partial \phi} \right)^2 - 2a \sin^2 \theta \frac{%
\left(1-F(r)\right)}{r^2 F(r)} \left( \frac{\partial S}{\partial t} \right)
\left( \frac{\partial S}{\partial \phi} \right) \\
& + F(r) \left( \frac{d S_r}{d r} \right)^2 +\frac{1}{r^2 \sin^2 \theta}
\left( \frac{d S_\theta}{d \theta} \right)^2 +\mathcal{O}\left( a^2 \right).
\end{align}
Additional computations and reductions yield
\begin{align}
r^4 F(r) \left( \frac{d S_r}{d r} \right)^2 &=\frac{E^2 r^4}{F(r)}+2a E L
r^2 \frac{\left(1-F(r)\right)}{F(r)}-r^2 \left(L^2 + \mathcal{K} \right), \\
\frac{1}{\sin^2 \theta} \left( \frac{d S_\theta}{d \theta} \right)^2 & =%
\mathcal{K}-2a E L \frac{\left(1-F(r)\right)}{F(r)} \cos^2 \theta.
\end{align}
where $\mathcal{K}$ represents the separation constant. We derive the whole
set of equations characterizing the photon motion using the definition of
the canonically conjugate momentum $p_\mu=g_{\mu \nu} \frac{dx^\nu}{d\tau}$.
In this way, we get
\begin{align}  \label{dphidtau}
r^2\frac{dt}{d \tau}&=\frac{E r^2}{F(r)} -a L \frac{\left( 1-F(r) \right)}{
F(r)}\sin^2\theta, \\
r^2\frac{dr}{d \tau}&=\sqrt{R(r)}, \\
r^2 \frac{d \theta}{d \tau}&=\sqrt{\Theta(\theta)}, \\
r^2\frac{d \phi}{d \tau}&=L+a E \frac{ \left(1-F(r)\right)}{F(r)}%
\sin^2\theta.
\end{align}
where the formulas of $R(r)$ and $\Theta(\theta)$ are
\begin{align}  \label{Rr}
R(r)&= E^2 r^4 +2aELr^2 \left( 1-F(r) \right)-r^2 F(r)(L^2+\mathcal{K}), \\
\Theta(\theta) &= \mathcal{K} \csc^2 \theta-2a E L \frac{\left(1-F(r) \right)%
}{F(r)} \cot^2 \theta.
\end{align}
An appropriate method for examining the geometric shapes of the shadow and
to derive the acoustic and photon sphere would be to take into account the
effective potential, which takes the following form
\begin{equation}  \label{Ve}
V_{eff}(r)=-\left( \frac{dr}{d \tau} \right)^2=-E^2+\frac{2aEL}{r^2}%
\left(F(r) - 1 \right) +\frac{F(r)}{r^2}\left( L^2+\mathcal{K} \right).
\end{equation}
It is worth remarking that the obtained potential matches the regular %
LTBH black hole when $\xi= 0$, i.e $%
F(r)=f(r)=1-\frac{2M}{r}$. We present the following impact factors in order to investigate the geometrical shape of the shadow in the slowly rotating regime
\begin{equation}
\zeta=\frac{L}{E}, \quad \eta=\frac{\mathcal{K}}{E^2}.
\end{equation}
These two impact parameters are used to rewrite the function $R(r)$ given in %
\eqref{Rr} as
\begin{equation}
R(r)=E^2 \left( r^4 +2a \zeta r^2\left( 1-F(r) \right)-r^2 \,
F(r)(\zeta^2+\eta) \right).  \label{fRr}
\end{equation}
The following conditions allow for the direct derivation of the critical
unstable circular orbits
\begin{equation}
R(r) \big\vert_{r_i}=0, \quad \frac{d R(r)}{dr}\big\vert_{r_i}=0.
\label{Cond}
\end{equation}
Thus, using \eqref{fRr} and \eqref{Cond}, we obtain the impact parameters $%
\eta$ and $\zeta$, which are
\begin{align}  \label{xi}
\zeta &=\frac{r \left(r F^{\prime }(r)-2 F(r)\right)}{2 a F^{\prime }(r)}, \\
\label{eta}
\eta & = \frac{r \left( 8 a^2 F^{\prime }(r)-\left(r F^{\prime }(r)-2
F(r)\right) \left(\left(r^2-4 a^2\right) F^{\prime }(r)-2 r F(r)\right)
\right)}{4 a^2 F^{\prime 2}}.
\end{align}
The shadow is then governed by the following equation
\begin{align}
R_s^2= \zeta^2+ \eta.  \label{Rshadow}
\end{align}

\subsection{Photon and acoustic spheres}

\begin{figure}[h]
\begin{tabular}{lll}
\includegraphics[scale=0.32]{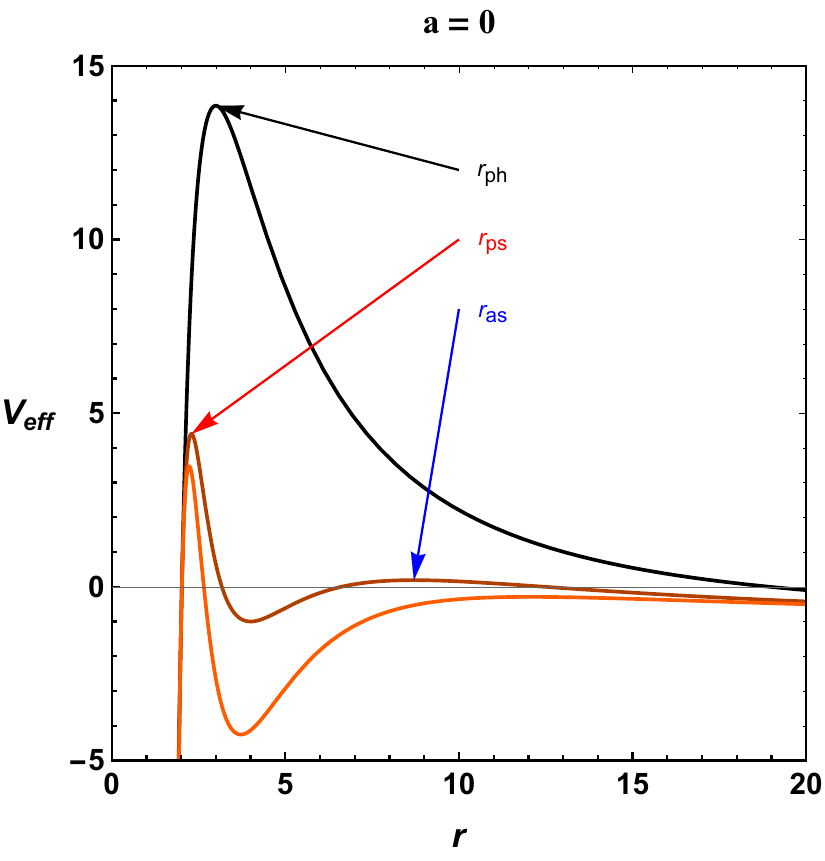} & \includegraphics[scale=0.32]{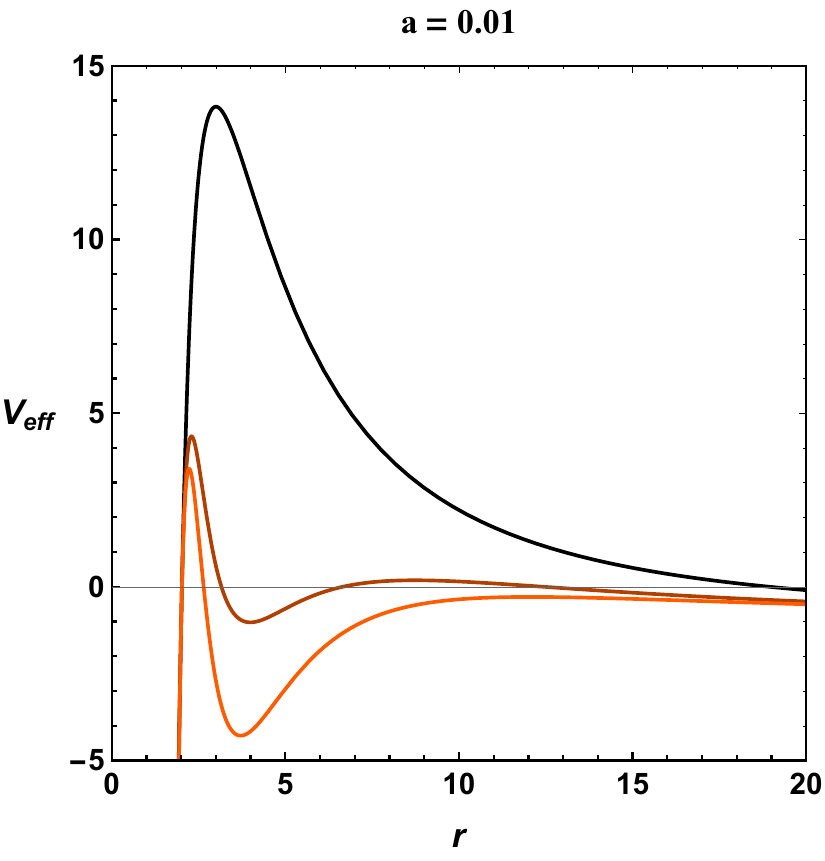} & %
\includegraphics[scale=0.32]{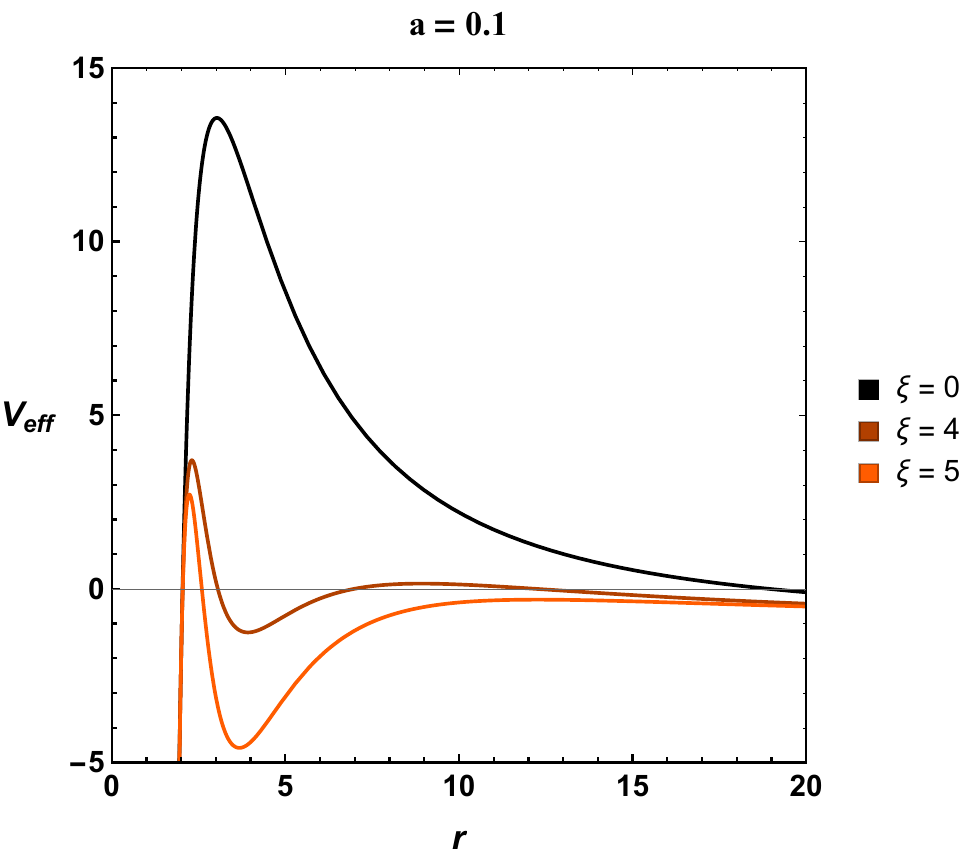}%
\end{tabular}%
\caption{{\protect\footnotesize Effective potential for different values of
the acoustic parameter $\protect\xi$ and different values of the rotating
parameter $a$. The black curve is associated to the LTBH
black hole. We take $M=1$, $E=\mathcal{K}=1$ and $L=20$.}}
\label{V}
\end{figure}

In this part of the paper, we use the equations established in the previous
section to inspect the different regions that can be obtained from the
effective potential of equation \eqref{Ve}. To examine its behaviors and determine its maximum values, we illustrate such quantity versus $r$ for
different values of the spin parameter $a$ and the acoustic parameter $\xi$ in
figure \eqref{V}.

\begin{center}
\begin{figure}[h]
\begin{tabular}{cc}
\hspace{1cm} \includegraphics[width=5.5cm,height=5.5cm]{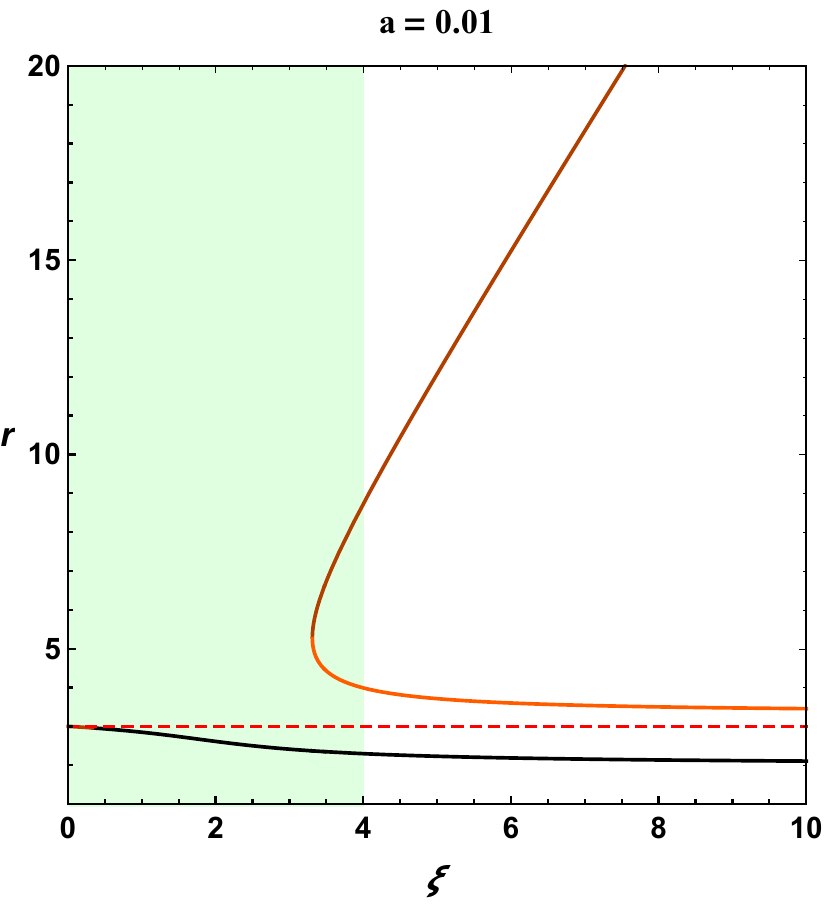} & \hspace{1cm} %
\includegraphics[width=6.5cm,height=5.5cm]{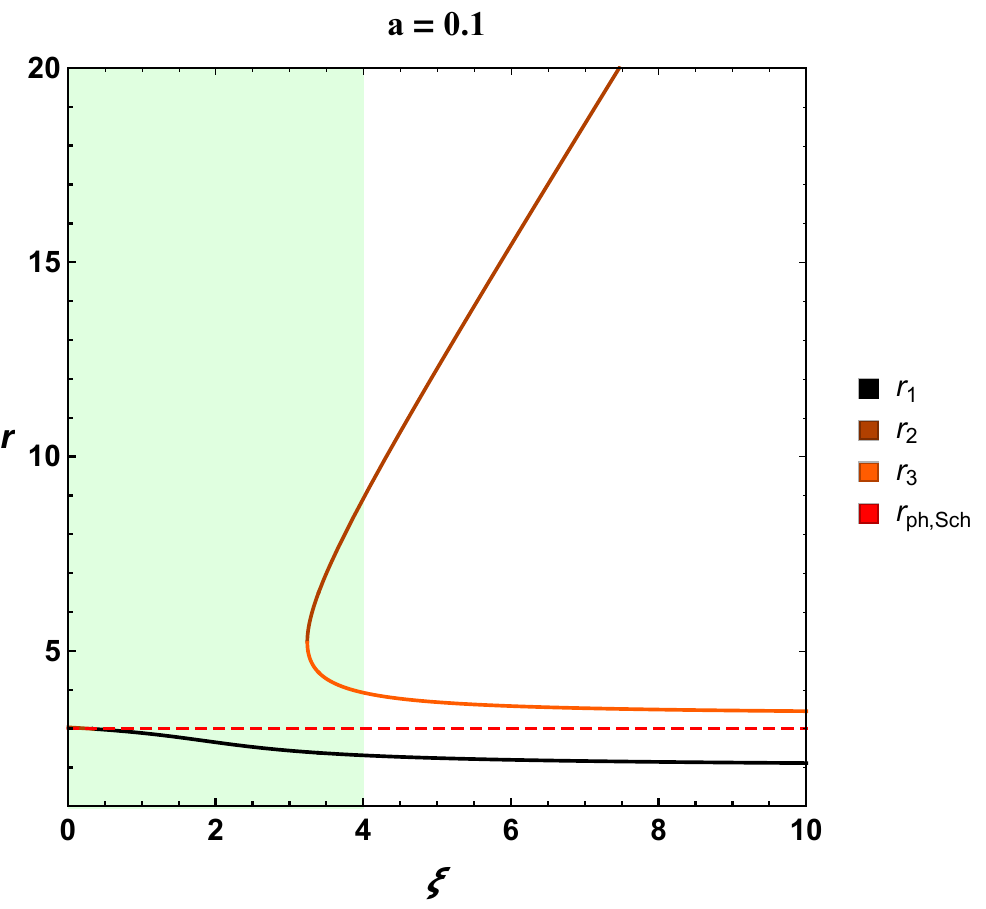}%
\end{tabular}%
\caption{ {\protect\footnotesize Roots of the effective potential derivative
against the acoustic parameter $\protect\xi$. The red dashed curve is
associated to photon sphere radius of the LTBH black hole. We take $M=1$, $E=%
\mathcal{K}=1$ and $L=20$.}}
\label{Vradii}
\end{figure}
\end{center}

From such a figure, we remark that the effective potential has different
extrema when the acoustic parameter $\xi \geq 4$. The most relevant are the
two maximum which are associated to the photon sphere radius $r_{ps}$ and to
the acoustic sphere $r_{as}$ while such a behavior vanishes when $\xi = 0$
where only one maximum associated to the photon sphere radius $r_{ph}$ is
present. Such radii are represented in the left panel of figure \eqref{V}
for the case of $a=0$. With these plots in hand, it can be observed that
the photon sphere radius $r_{ps}$ decreases when higher values of $\xi$ are
considered, while the opposite behavior is observed for the acoustic sphere
radius $r_{as}$. From the three different panels, we can deduce that the
higher the rotation parameter $a$ gets, the larger these radii are. To
investigate the extrema of the effective potential, we derive the later and
solve the following equation
\begin{equation}
V_{eff}^{\prime}(r)\vert_{r=r_i}=\frac{4 a E L-\left(2 F(r)-r F^{\prime
}(r)\right) \left(2 a E L+\mathcal{K}+L^2\right)}{r^3}=0.
\end{equation}
By defining $\left(2 a E L+\mathcal{K}+L^2\right)=A$, $4 a E L=B$ and
expliciting the factor $\left(2 F(r)-r F^{\prime }(r)\right)$, we obtain
\begin{equation}
\frac{A \left(40 M^3 \xi -32 M^2 \xi r+6 M (\xi +1) r^2-2 r^3\right)+B r^3}{%
r^6}=\frac{-\textcolor{blue}{e} r^3+b r^2-c r+d}{r^6}=0,
\end{equation}
where $\textcolor{blue}{e}=2 A-B, b=6 A M (\xi +1), c=32 A M^2 \xi, d=+40 A
M^3$. It can be shown that such an equation admits three solutions that are
\begin{align}
r_1 &=\frac{b}{3 e}+\frac{2^{1/3} \left(3 e c-b^2\right)}{3 e \left( X
\right)^{1/3}}-\frac{\left( X \right)^{1/3}}{3 \sqrt[3]{2} e}, \\
r_2 &=\frac{b}{3 e}-\frac{\left(1+i \sqrt{3}\right) \left(3 e c-b^2\right)}{%
3\ 2^{2/3} e \left( X \right)^{1/3}}+\frac{\left(1-i \sqrt{3}\right) \left(
X \right)^{1/3}}{6 \, 2^{1/3} \, e}, \\
r_3 &=\frac{b}{3 e}-\frac{\left(1-i \sqrt{3}\right) \left(3 e c-b^2\right)}{%
3\ 2^{2/3} e \left( X \right)^{1/3}}+\frac{\left(1+i \sqrt{3}\right) \left(
X \right)^{1/3}}{6 \, 2^{1/3} \, e},
\end{align}
where
\begin{equation}
X=\sqrt{\left(27 e^2 d-9 e b c+2 b^3\right)^2-4 \left(b^2-3 e c\right)^3}-27
e^2 d+9 e b c-2 b^3.
\end{equation}
To fully identify these quantities further investigation should be carried. Indeed, to visualise the critical radii behaviors, we plot the three roots of the
effective potential derivative against the acoustic parameter $\xi$ in figure %
\eqref{Vradii}.
With these figures and after a numerical analysis, it appears that when $\xi=0$ the
solution $r_2$ can be associated to the photon sphere radius $r_{ph}$ while
for $\xi \geq 4$ such a solution is associated to the acoustic sphere radius
$r_{as}$.

To analyse each radius in detail, we first compute the effective potential and its derivative for a LTBH $(\xi=0)$
\begin{align}
V_{eff} &=-\frac{4 a E L M+L^2 (2 M-r)+2 K M+E^2 r^3-K r}{r^3}, \\
V_{eff}^{\prime} &= \frac{12 a E L M+L^2 (6 M-2 r)+6 K M-2 K r}{r^4}.
\end{align}
The solution to $V_{eff}^{\prime}=0$, is given by $r_{ph}=M \left(\frac{6 a E L}{K+L^2}%
+3\right)$. For the case where $a=0$, it matches the Schwarzschild
photon sphere $r_{ph}=3M$. In this way, the LTABH can be linked to the Schwarzschild black hole by turning off the acoustic parameter $\xi$ and the rotation parameter $a$ which results as follows $%
r_1=0 $, $r_2=r_{ph}=3M$, $r_3=0$. At the same time, it can be seen from the plots that the radius $%
r_2$ plays another role when the value of the acoustic parameter $\xi \geq 0$. Actually, it is associated to the acoustic sphere
radius $r_{as}$ since its values are expected to be the largest. Seeing that $r_3$ corresponds to the minimum of the effective
potential, such a radius will not be considered because it is not physically
relevant. Finally, we have the solution $r_1$ which represents the photon sphere
radius $r_{ps}$ for $\xi \geq 4$ and have the lowest values compared to the
other critical radii. Concerning the impact of the parameter $a$, its effect
is barely visible.

From such an analysis, we expect the LTABH to be
characterized by two shadows. The first, is the \textit{optical shadow} $R_s\left(
r_{ps} \right)$ associated with the behavior of photons near the horizon
while the second is the \textit{acoustic shadow} $R_s\left( r_{as} \right)$ that
describes the characteristics of sound waves approaching the acoustic
horizon.

\subsection{Shadow forms}

\begin{figure}[h]
\begin{tabular}{ll}
\includegraphics[scale=0.45]{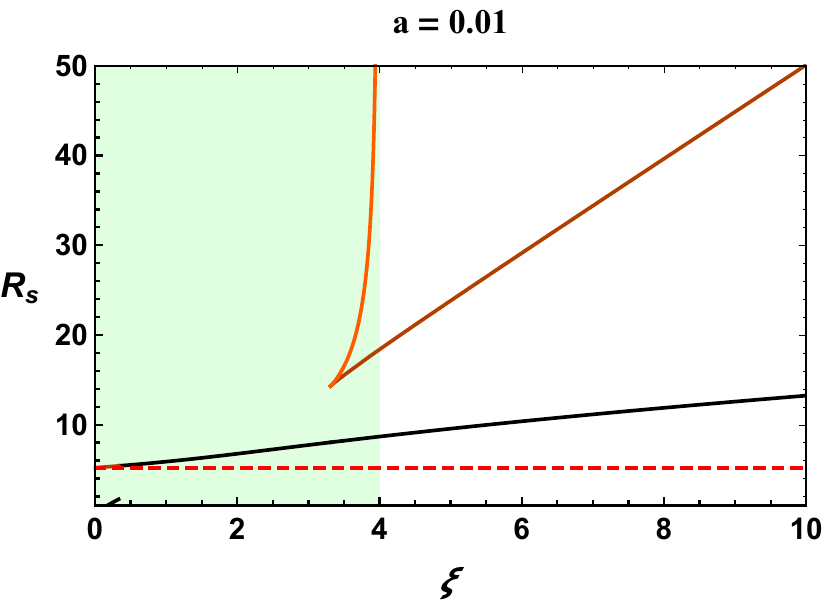} & \includegraphics[scale=0.45]{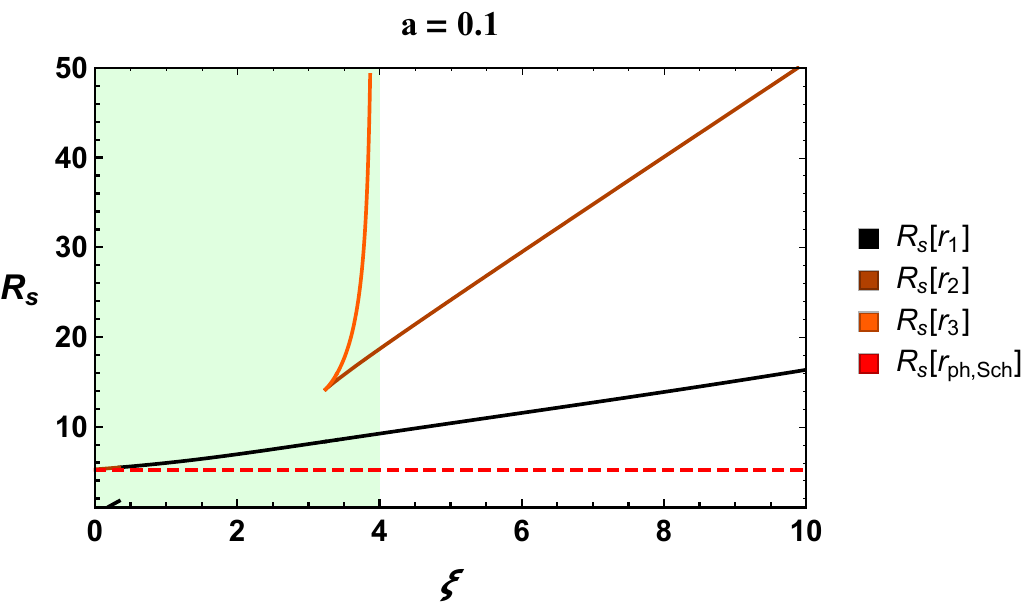}%
\end{tabular}%
\caption{ {\protect\footnotesize Shadow radius as a function of the
effective potential extrema against the acoustic parameter $\protect\xi$. The
red dashed curve is associated to photon sphere radius of the
LTBH black hole for which $R_s=3\protect\sqrt{3}M$. We
take $M=1$, $E=\mathcal{K}=1$ and $L=20$.}}
\label{PShad}
\end{figure}

As we have previously mentioned, in the frame of a LTABH, the optical shadow surrounding the event horizon, which
describes the visual boundary that light cannot escape from by viewers, and
the acoustic shadow, which describes the audible boundary of the sound waves
detected by static listeners, may coexist. In what follows, we derive the
equation governing such aspects of the black hole where the photon-phonon
interaction has been neglected to avoid complications. Using equations %
\eqref{xi}, \eqref{eta} and
considering that the observer
is located at the equatorial plan $\left( \theta=\frac{\pi}{2} \right)$, we
illustrate the shadow radius denoted by $R_s$ as a function of the acoustic
parameter $\xi$ in figure \eqref{PShad} for the different radii $r_i$
derived from the previous section, with $i=1,2,3$.

As expected, the behavior of the shadow radius $R_s (r_3)$ is not physically
relevant since its values range in the green region where the values of the
acoustic parameter $\xi$ are not allowed. We also observe from such a figure
that the shadow radius of the LTBH black hole coincide with $R_s(r_2)$ for
the particular value of $\xi=0$ proving that the root $r_2$ can be
associated with the photon sphere in such a case. When higher values of the
acoustic parameter $\xi$ are taken into consideration, we have two shadows.
The acoustic shadow $R_s\left( r_{as} \right) := R_s \left( r_{2} \right)$
and the usual shadow $R_s\left( r_{ps} \right) := R_s \left( r_{1} \right)$.
An interesting behavior arises from such a figure: indeed, we remark that
increasing the rotating parameter $a$ has a bigger impact on the shadow
radius than the Acoustic shadow radius. This impact is also more visible for
higher values of $\xi$.

\begin{figure}[H]
\begin{tabular}{ll}
\hspace{0cm} \includegraphics[scale=0.41]{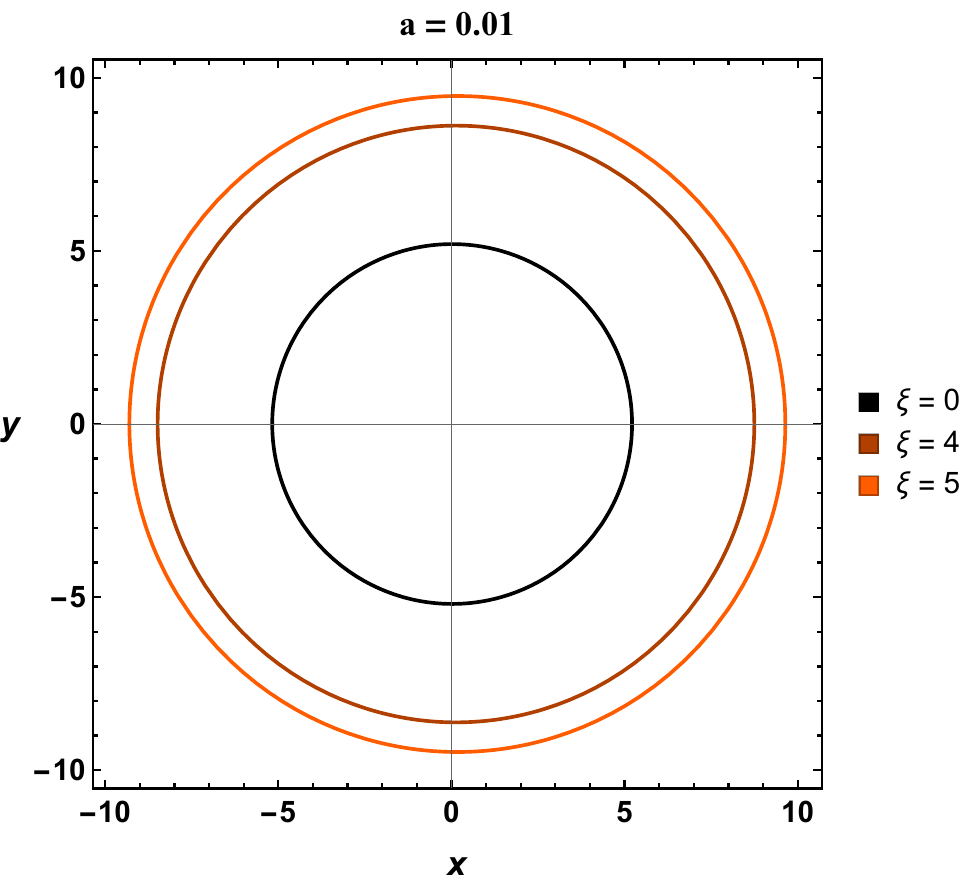} & %
\includegraphics[scale=0.41]{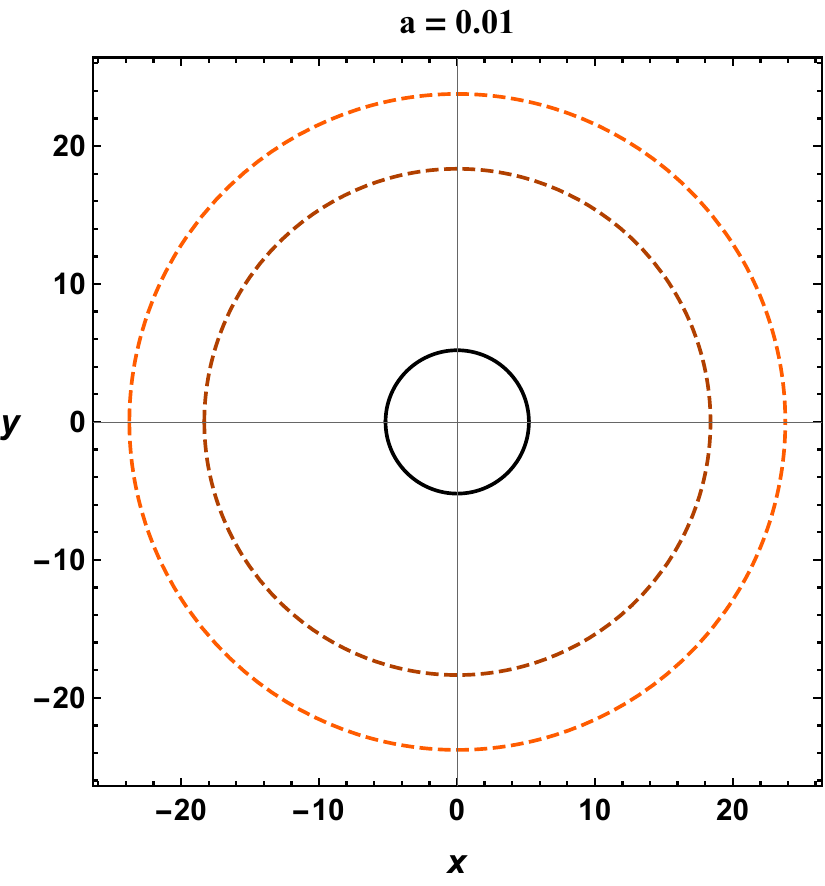} \\
\hspace{0cm} \includegraphics[scale=0.375]{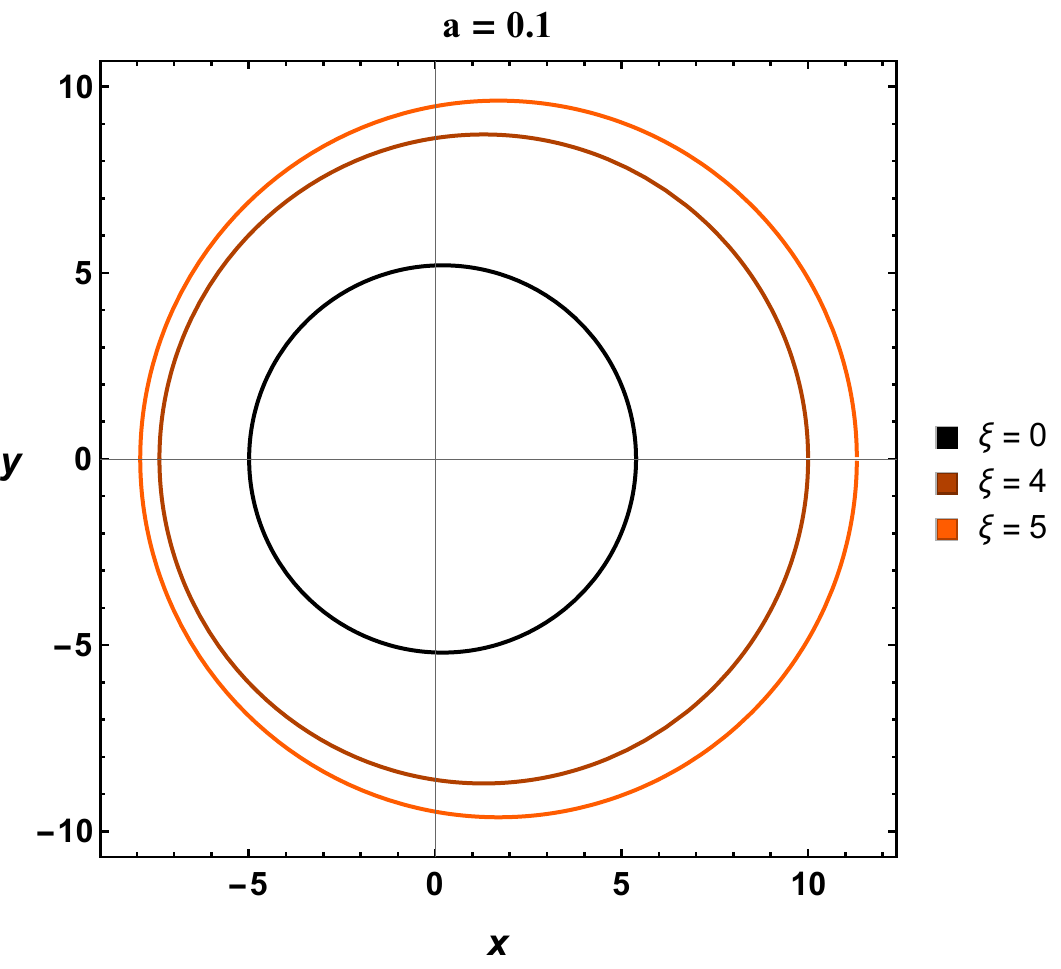} & %
\includegraphics[scale=0.41]{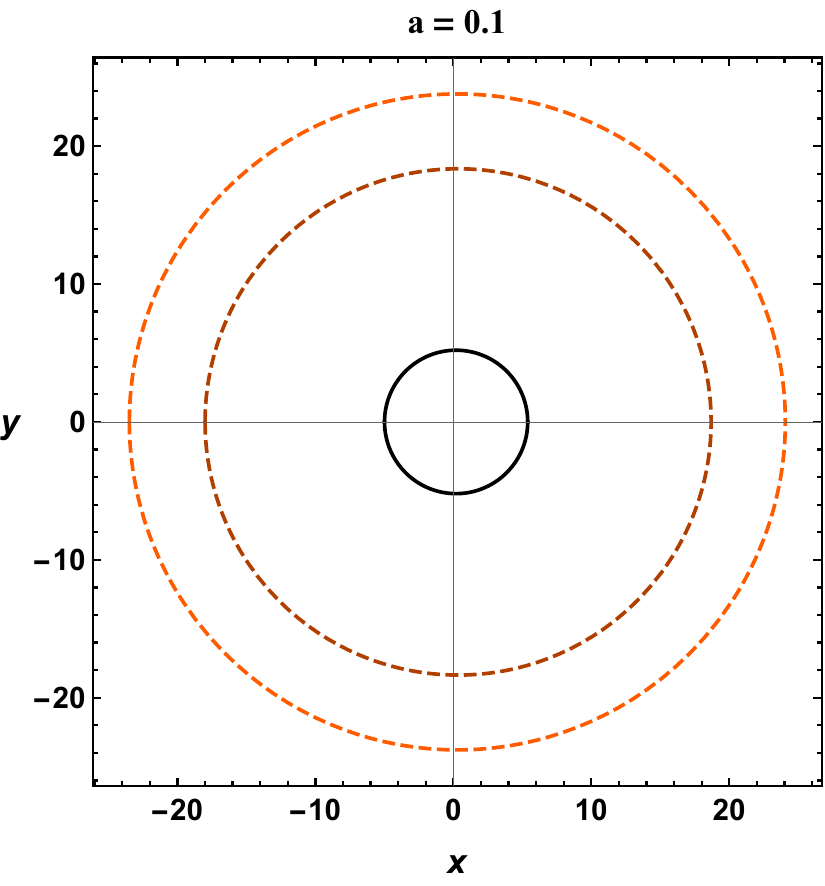}%
\end{tabular}%
\caption{{\protect\footnotesize Left: Regular shadow $R_s\left( r_{ps}
\right)$. Right: Acoustic shadow $R_s\left( r_{as} \right)$. We consider for
these illustrations different values of the acoustic parameter $\protect\xi$, the spin $a$ and a fixed black hole mass $\left( M=1 \right)$. }}
\label{Shad}
\end{figure}

The shadow geometrical shape is regulated by the permitted values of $\zeta$
and $\eta$. But a more precise method to visualize the shadow from a
distance would be to utilise the celestial coordinates $x$ and $y$
defined by
\begin{align}
x&= \lim_{r_* \to \infty} \left( -r^2_* \sin^2 \theta_0 \frac{d \phi}{dr}
\right), \\
y&= \lim_{r_* \to \infty} r^2_* \frac{d \theta}{dr},
\end{align}
where $r_*$ is the distance between the black hole and the observer and $%
\theta_0$ is associated to the inclination angle between the line rotational
axis of the black hole and the observer line of sight \cite{2S1}. As a
function of the impact parameters, these two celestial coordinates can be
written as
\begin{align}  \label{cel2}
x&= -\zeta \csc \theta_0, \\
y&= \sqrt{\eta -\zeta^2 \cot^2 \theta_0},
\end{align}
With the use of the celestial coordinates, we illustrate the regular and acoustic shadow at the equatorial plan in figure \eqref{Shad} for different values of $\xi$ and $a$.

From such a figure, interesting features arises. Indeed, we observe that for
low values of the rotating parameter $a$, both shadows are circular. In this
case, increasing the value of the acoustic parameter $\xi$ is associated to
larger shadows. When the rotating parameter $a$ is increased, we remark that
the acoustic shadow (right plots) maintains the circular shape which proves
that the acoustic shadow is barely affected by the rotation of the black
hole. Concerning the regular shadow, the frame dragging effect shifts the
shadows to the right side when $a$ is increased. Such illustrations show
that we must expect larger shadows in the case of acoustic black hole
compared to other known black hole like Kerr, Reissner-Nördstrom or
Schwarzschild.

To complete the essential building blocks associated with the optical
behavior of this solution, an examination of the shadow size and shape is
imposed. By examining these aspects, we can understand the geometric shadow
behaviors, whether the regular or the acoustic one. This analysis is essential for
a comprehensive overview of the underlying physical phenomena and the
specific characteristics of the massive objects that generate these shadows,
such as LTABH. To examine the geometric deformations of the two shadow
types, the visual and acoustic one, two essential parameters are generally
used : ($R_s$, $R_{as}$) and ($\delta_s$, $\delta_{as}$) representing the
size and the deformation of visual and acoustic shadows, respectively. The
size of the shadow is specifically defined by three key points: the upper
position ($x_t,y_t$), the bottom position ($x_b, y_b$) and the circle
reference point ($x_r,0$) associated with the non deformed geometry. The
point where the deformed shadow circle intersects the horizontal axis is
denoted ($x_i,0$). The distance between the reference point and the point of
intersection, provides a quantifiable measure of the geometric deviation
caused by the variation of the relevant parameters controlling the present
solution. Consequently, the shadow radius $R_{s/as}$ and the deformation $%
\delta_{s/as}$ are approximately determined by the following expressions
\begin{equation}  \label{ }
R_{s/as}=\frac{(x_t-x_r)^2+y_t^2}{x_t-x_r}
\end{equation}
and
\begin{equation}  \label{ }
\delta_{s/as}=\frac{(x_i-x_p)}{R_{s/as}}
\end{equation}
where $x_r$ is the rightmost position of the circle. Based on these
expressions, we can calculate the size and the geometrical deformation of
the visual and acoustic shadows. Indeed, in table.(\ref{acc}), we report these quantities with respect to the spin and the acoustic parameter variations. A close
examination show that the radius of the regular and the acoustic shadow increase
with $\xi$ for fixed values of the spin parameter. By fixing $\xi$, we can
distinguish between the spin impact on the shadow radius and the acoustic
radius. In fact, we observe that the quantity $R_s$ increases while the
quantity $R_{as}$ undergoes a negligible change. Moreover, the acoustic shadow
is bigger than the regular one for fixed values of the parameters $a$ and $%
\xi$. This particularity could be linked to the fundamental differences in size
between the radius of the photon sphere and the region of acoustic shadow
formation. This means that sound waves can bypass objects more easily.
Regarding the distortion parameter $\delta_{s/as}$, it is almost constant by varying $\xi$. By increasing such a
parameter, we notice that the distortion decreases. In this regard, we
remark a small difference in the distortion parameter by varying the spin.
Besides, it can be observed that the spin has more impact on $\delta_{s}$
than $\delta_{as}$.
\begin{table}[H]
\begin{center}
\scalebox{0.85}{
\begin{tabular}{|c|ccc|ccc|}
\hline
\multirow{2}{*}{} & \multicolumn{3}{c|}{$a=0.01$}                                                                   & \multicolumn{3}{c|}{$a=0.1$}                                                                     \\ \cline{2-7}
                  & \multicolumn{1}{c|}{$\xi=0$}         & \multicolumn{1}{c|}{$\xi=4$}         & $\xi=5$        & \multicolumn{1}{c|}{$\xi=0$}         & \multicolumn{1}{c|}{$\xi=4$}         & $\xi=5$         \\ \hline
$R_s$             & \multicolumn{1}{c|}{$5.19$}           & \multicolumn{1}{c|}{$8.61$}           & $9.47$          & \multicolumn{1}{c|}{$5.19$}           & \multicolumn{1}{c|}{$8.71$}           & $9.62$           \\ \hline
$\delta_s$        & \multicolumn{1}{c|}{$1.45\times 10^{-4}$} & \multicolumn{1}{c|}{$4.08\times10^{-5}$} & $2.7\times 10^{-4}$ & \multicolumn{1}{c|}{$2.23\times 10^{-3}$} & \multicolumn{1}{c|}{$1.78\times 10^{-3}$} & $1.13\times10^{-3}$ \\ \hline
$R_{as}$            & \multicolumn{1}{c|}{$5.19$}           & \multicolumn{1}{c|}{$18.35$}          & $23.78$         & \multicolumn{1}{c|}{$5.19$}           & \multicolumn{1}{c|}{$18.35$}          & $23.78$          \\ \hline
$\delta_{as}$       & \multicolumn{1}{c|}{$1.45\times 10^{-4}$} & \multicolumn{1}{c|}{$1.89\times 10^{-5}$} & $1.9\times 10^{-5}$ & \multicolumn{1}{c|}{$2.23\times 10^{-3}$} & \multicolumn{1}{c|}{$6.86\times 10^{-4}$} & $2.56\times 10^{-4}$ \\ \hline
\end{tabular}}
\end{center}
\caption{Geometrical deformation of slowly rotating acoustic black hole by
varying the spin and $\protect\xi$ parameters.}
\label{acc}
\end{table}

\section{Lense-Thirring effect and precession}

\label{v}

In this part of the paper, we study the frame-dragging effect (i.e.,
Lense-Thirring precession) of the LTABH on the spin of a test gyroscope. One
of the key predictions of Einstein's general relativity is the
Lense-Thirring precession, which has historically been the subject of
enormous research efforts \cite{body1, body2}. The spin precession of a test
gyroscope in the gravitational field of a spinning spherical body was first
reported by Josef Lense and Hans Thirring in 1918. The Lense-Thirring
precession effect can be used to anticipate the causal structures of the
spacetime geometries. Consequently, it is crucial to understand the
Lense-Thirring precession phenomenon in order to differentiate between the
different black holes. In the following section, we review the mathematical
description of the Lense-Thirring effect \cite{bodyT}. Then, we use the
findings to analyse the frame-dragging in the vicinity of a LTABH.

\subsection{Mathematical Framework}

In the what follows, we examine a straightforward situation in which the
gyroscopes are fixed at each point in space, and we look into how the
precession frequency of the test gyroscope's spin varies as the spatial
points move. The gyroscope moves along an integral curve $\gamma(t)$ of a
timelike Killing vector field $K$ since it is at rest in a stationary
spacetime \cite{body3}. Then, the gyroscope's four-velocity $u^\alpha$ can
be expressed in terms of $K^\alpha$ as follows
\begin{equation}
u^\alpha=\frac{1}{\sqrt{-K^\beta K_\beta}}K^\alpha.
\end{equation}
In this way, one could express the four-velocity vector of the timelike
curve $\gamma(\tau)$ as $u=\dot{\gamma}$, with $u$ being tangent to $%
\gamma(\tau)$ and satisfying the condition $u^\mu u_\mu=-1$ and $\tau$ being
the proper time. It is worth noting that the Fermi derivative component of
any vector in the $u$-direction can be written as
\begin{equation}
u^\alpha F_\alpha X^\beta=u^\alpha \nabla_\alpha X^\beta-X^\alpha a_\alpha
u^\beta +X^\alpha u_\alpha a^\beta,  \label{Eq1}
\end{equation}
where the acceleration is $a^\alpha=u^\beta \nabla_\beta u^\alpha$. Given
the definition $e^\alpha_\mu e_{\nu \alpha}=\eta_{\mu
\nu}=diag \left( -1,1,1,1 \right)$ in which the orthonormal frame $%
\left\lbrace e_i \right\rbrace \, \, \left(i=1,2,3\right)$ is perpendicular
to $e_0=u$, we may write the components of the co-variant derivative in the
direction of four-velocity $u_{\alpha}$ of the orthonormal basis as
\begin{equation}
u^\beta \nabla_\beta e_i^\alpha = e_i^\beta a_\beta u^\alpha+\omega_i^j
e_j^\alpha,  \label{Eq2}
\end{equation}
with $\omega_i^j=u^\alpha \nabla_\alpha e^j_\beta e^\beta_i$. With the help
of the equations \eqref{Eq1} and \eqref{Eq2}, the Fermi derivative component
of the orthonormal basis in the direction of four-velocity $u_{\alpha}$ can be
expressed as
\begin{equation}
u^\beta F_\beta e_\alpha = \omega_\alpha^\beta e_\beta.
\end{equation}
It is worth noticing that $u^\beta F_\beta e_\alpha \neq 0$ for a rotating
spacetime only. On another hand, let us note $\mathcal{S}$ as the spin
vector of a rigid rotator, such as a gyroscope, or the expected value of the
spin operator for a particle. The change in spin $\mathcal{S}$ with respect
to proper time $\tau$ is expressed after expanding $u^\beta F_\beta \mathcal{%
S} = 0$ as follows
\begin{equation}
\frac{d\mathcal{S}^\alpha}{d\tau}=\omega_\beta^\alpha \mathcal{S}^\beta,
\end{equation}
suggesting that the spin vector $\mathcal{S}$ undergo precession with a
certain frequency. To derive the needed quantity, we introduce the angular
frequency
\begin{equation}
\omega_{ij}=\epsilon_{ijk} \omega^k,
\end{equation}
where $\epsilon_{ijk}$ is the levi-civita tensor and $\omega^k$ represents
the Lense-Thirring precession frequency. The angular velocity is described
in terms of timelike Killing vector field $K^{\alpha}$ by the equation
\begin{equation}
\omega_{ij}=\frac{1}{\sqrt{-K^\alpha K_\alpha}} e^\beta_j K^\delta
\nabla_\delta e_{i \beta}.
\end{equation}
As a result the quantity $\Omega$ is given by
\begin{equation}
\Omega_{\mu}= \frac{1}{2 \sqrt{-K^\alpha K_\alpha}} \ast \left( \overline{K}
\wedge d\overline{K} \right)_{\mu},
\end{equation}
where $\wedge$ is the external product and $\ast$ is the Hodge operator \cite%
{body4}. The Lense-Thirring precession frequency for the spin vector $%
\mathcal{S}$ can be expressed as follows
\begin{equation}
\Omega_\mu= \frac{1}{2 \sqrt{-K^\alpha K_\alpha}} \mathcal{N}_\mu^{\nu \rho
\sigma} K_\nu \partial_\rho K_\sigma,  \label{O1}
\end{equation}
where the canonical volume associated to the metric $g_{\mu \nu}$ is $%
\mathcal{N}$ and $\mathcal{N}_\mu ^{\nu \rho \sigma}=\frac{1}{4 \!} \sqrt{-g}
g_{\mu \alpha} \epsilon^{\alpha \nu \rho \sigma}$ with $\epsilon^{\alpha \nu
\rho \sigma}$ being Levi-Civita symbol . The covariant form of the timelike
Killing vector $K^\mu$ is $K_\mu=g_{\mu \nu}K^\nu$. In the case of a
stationary spacetime, we can write $K^\mu=\left(1,0,0,0 \right)$ and $%
K_\mu=g_{\mu t}$ which yields the decomposition $K_\mu dx^\mu = g_{tt} dt +
g_{ti} dx^i$ for a stationary test gyroscope. In this way, for a spacetime
described by a general slowly rotating metric
\begin{equation}
ds^2=-g_{tt} dt^2 + g_{rr}dr^2+g_{\theta \theta} d\theta^2+ g_{\phi \phi}
d\phi^2 -2 g_{t \phi} dt d\phi.
\end{equation}
we obtain the following magnitude of the spin precession frequency
\begin{equation}
\Omega= \left( g_{\theta \theta} \left( \Omega_{\theta} \right)^2 + g_{r r}
\left( \Omega_{r} \right)^2 \right)^{\frac{1}{2}},  \label{O2}
\end{equation}
where one has used the following equations \cite{bodyT}
\begin{equation}
\Omega_{\theta} = \frac{g_{tt}}{2\sqrt{-g}} \left( \frac{g_{t\phi}}{g_{tt}}
\right)_{,r}, \quad \Omega_{r} = \frac{g_{tt}}{2\sqrt{-g}} \left( \frac{%
g_{t\phi}}{g_{tt}} \right)_{,\theta}.  \label{O3}
\end{equation}

\subsection{Graphics and results}

\begin{figure}[h]
\begin{tabular}{lll}
\includegraphics[scale=0.39]{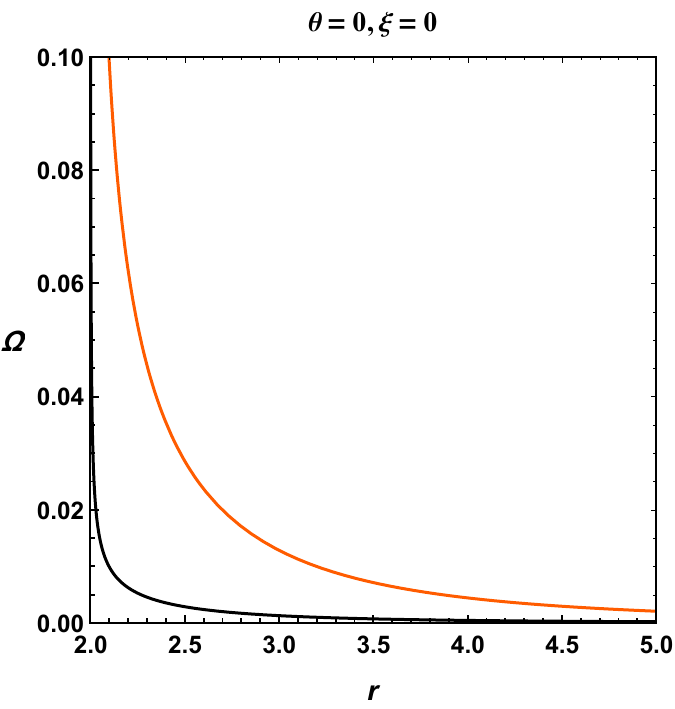} & \includegraphics[scale=0.39]{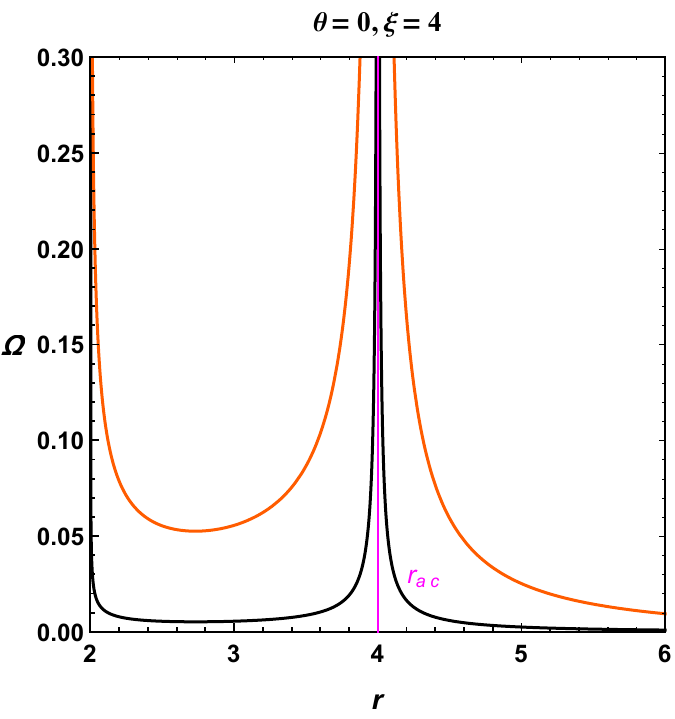} & %
\includegraphics[scale=0.39]{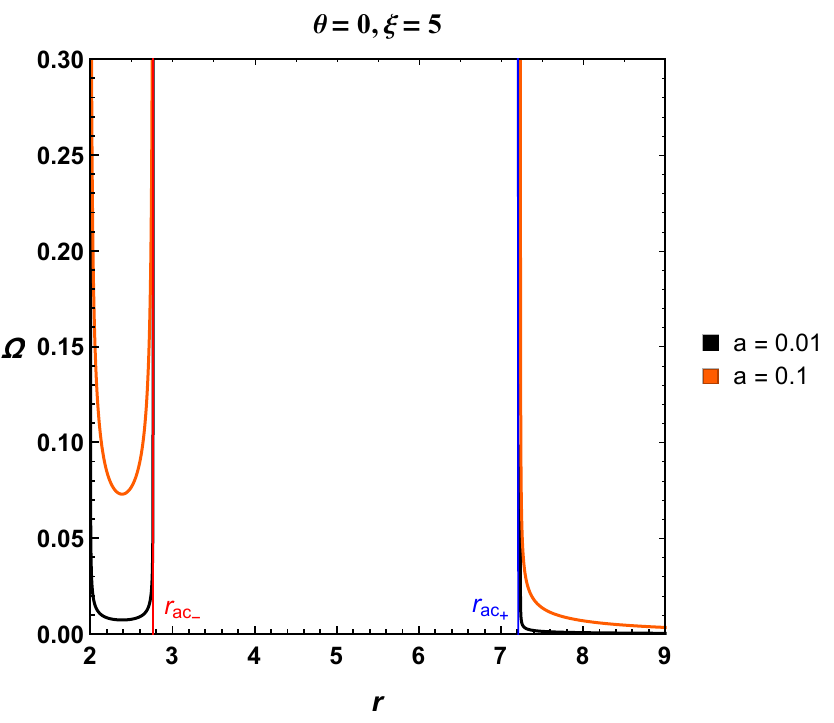} \\
\includegraphics[scale=0.39]{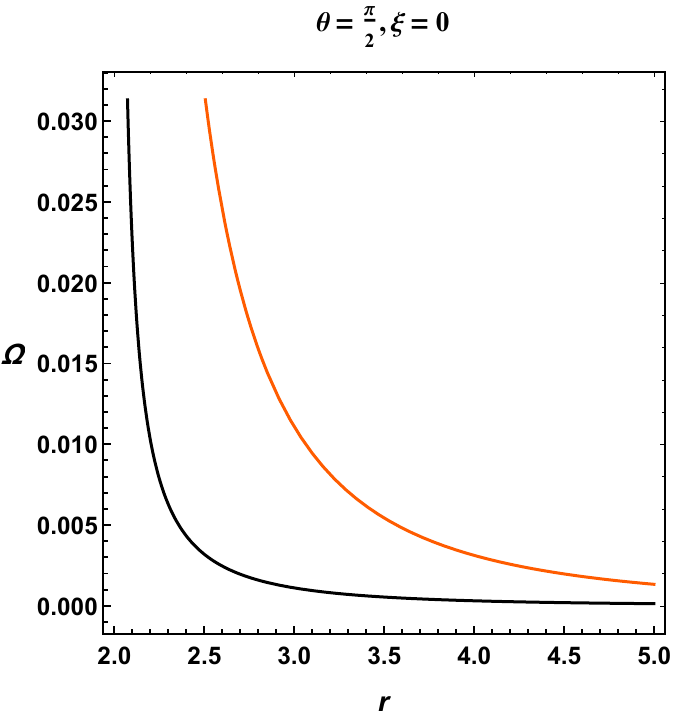} & \includegraphics[scale=0.39]{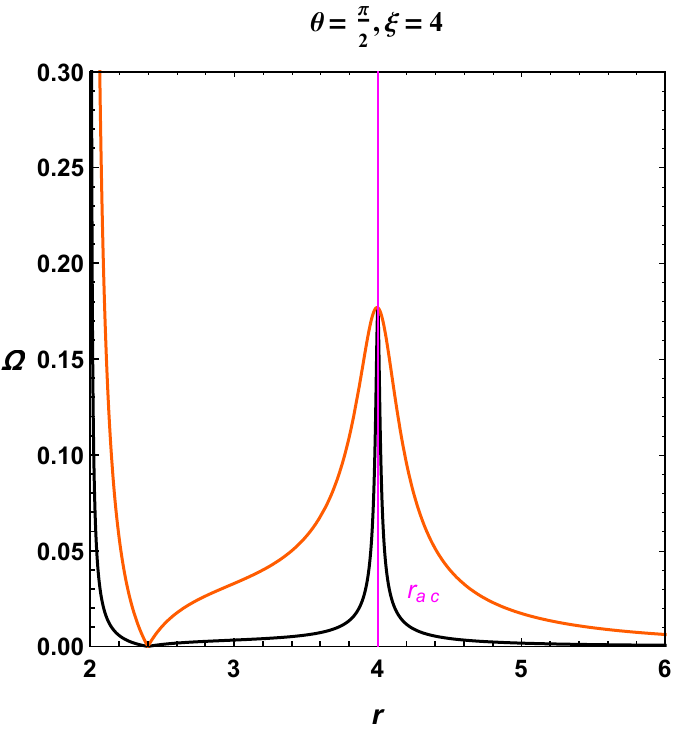} & %
\includegraphics[scale=0.39]{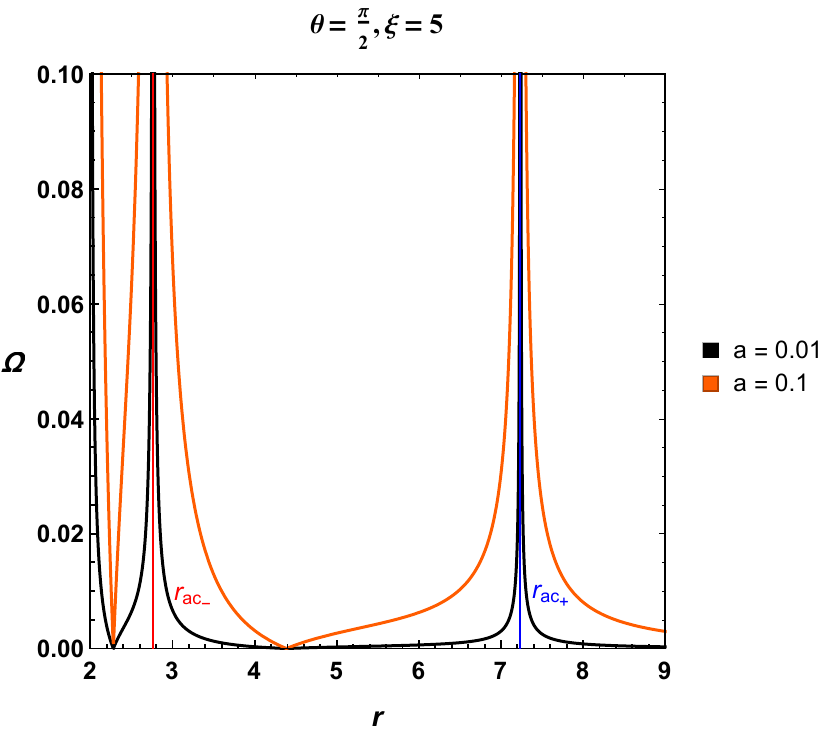}%
\end{tabular}%
\caption{ {\protect\footnotesize Lense Thirring precession in a slowly
rotating black hole for fixed $M=1$. Left: Non acoustic case. Middel:
Extremal case of the acoustic black hole. Right: Acoustic black hole. We
take $\protect\theta=0^{\circ}$ for the upper panels and $\protect\theta%
=90^{\circ}$ for the lower panels. }}
\label{LT}
\end{figure}

In this part of the paper, we use the obtained results and apply them to our
case which correspond to the acoustic black hole in the slowly rotating
regime. Indeed, with the use of equations \eqref{last}, \eqref{O2} and %
\eqref{O3}, we find
\begin{align}
\Omega_{\theta} & = \frac{a M \sin (\theta ) \left(\xi (r-6 M) (r-2
M)+r^2\right)}{r (2 M-r) \left(r^2-2 M \xi (r-2 M)\right) \sqrt{\Gamma}}, \\
\Omega_{r} & = \frac{a M \sin (2 \theta ) \left(\xi (r-2 M)^2+r^2\right)}{%
r^3 \sin (\theta ) \sqrt{\Gamma}},
\end{align}
where
\begin{equation}
\Gamma= r^4 -\frac{4 a^2 M^2 \sin ^2(\theta ) \left(\xi (r-2
M)^2+r^2\right)^2}{r (2 M-r) \left(r^2-2 M \xi (r-2 M)\right)}.
\end{equation}
The final magnitude $\Omega$ is given by
\begin{align}
\Omega &= \left[ -\frac{ a^2 M^2 }{2 r^2 (2 M-r) \left( \chi \right) \left(2
a^2 M^2 (\cos (2 \theta )-1) \left(\xi (r-2 M)^2+r^2\right)^2+r^5 (2 M-r)
\left( \chi \right)\right)} \right]^{\frac{1}{2}} \times  \notag \\
&\left[ \left(8 M \xi ^3 (r-2 M)^6+\xi ^2 r^2 (6 M-r) (r-2 M)^2 \left(16
M^2-22 M r+5 r^2\right) \right. \right.  \notag \\
& +2 \xi r^4 (2 M-r) (4 M-r) (6 M-5 r) +r^6 (8 M-5 r)  \notag \\
& +\cos (2 \theta ) \left(8 M \xi ^3 (r-2 M)^6+\xi ^2 r^2 (6 M-r) (r-2 M)^2
\left(16 M^2-10 M r+3 r^2\right) \right.  \notag \\
& \left. \left. +2 \xi \, r^4 (2 M-r) \left(24 M^2-14 M r+3 r^2\right) +r^6
(8 M-3 r)\right) \right]^{\frac{1}{2}},  \label{om}
\end{align}
where $\chi = r^2-2 M \xi (r-2 M)$. At the limit $\xi \to 0$, we obtain the
following magnitude
\begin{equation}
\Omega= \sqrt{-\frac{a^2 M^2 (\cos (2 \theta ) (8 M-3 r)+8 M-5 r)}{2 \left(
r^2 (2 M-r) \left(2 a^2 M^2( \cos (2 \theta ) -1 )+r^3 (2 M-r)\right) \right)%
}}.
\end{equation}
This result matches perfectly the Kerr magnitude after some trigonometric
rearrangements \cite{bodyT}. To reveal the behavior of such a quantity, we
illustrate in figure \eqref{LT} its behavior as a function of $r$ for
different values of the acoustic parameter $\xi$ and the rotation $a$.
From such a figure, we observe that the Kerr black hole magnitude is
recovered when the case $\xi=0$ is considered. In this case, the angular
velocity decreases with the radius $r$ since one moves away from the black
hole. Near the event horizon, the spacetime is being dragged around the
black hole causing the frame dragging effect to become extremely strong
which leads to a very high value of the Lense Thirring precession. For the
cases where $\xi \geq 4$, we remark that the value of $\Omega$ increases
significantly around the acoustic radius $r_{ac}$ for the extremal case, and
around the inner and outer acoustic horizon in which $\xi > 4$. Such a
result suggests that the frame-dragging effect is very strong in these
regions. This could be interpreted by the twist of the spacetime that can be
caused by the acoustic parameter $\xi$. Another important result arises from
such illustrations. Precisely, at the equatorial plan $\left(\theta=\pi/2
\right)$, we have regions in which the the Lense-Thirring precession is
null. These regions may be localised with the radial coordinate and are
given by
\begin{align}  \label{O01}
r_{{\Omega=0}}^- &=\frac{2 M \left(2 \xi -\sqrt{(\xi -3) \xi }\right)}{\xi +1%
}, \\
r_{{\Omega=0}}^+ &=\frac{2 M \left(2 \xi +\sqrt{(\xi -3) \xi }\right)}{\xi +1%
}.  \label{O02}
\end{align}
From the expression of the $\Omega$ given in equation \eqref{om}, it can be
seen that the angular velocity is null if $a=0$. Such a case suggests that
the Lense-Thirring effect is null for non rotating black hole. In our case,
we have found that similar scenarios exist in the LTABH for values of the
spin different than zero. Indeed, equations \eqref{O01}-\eqref{O02} show
regions where the spacetime is not dragged by the black hole rotation
parameter. This means that the particles orbits would not be subject of the
precession occurring due to the frame dragging effect \cite{body5}. Such a result is
essentially due to the acoustic parameter that cancels the rotating effect.

\section{Deflection angle}

\label{vi} 
\begin{figure}[]
\begin{tabular}{lll}
\includegraphics[scale=0.41]{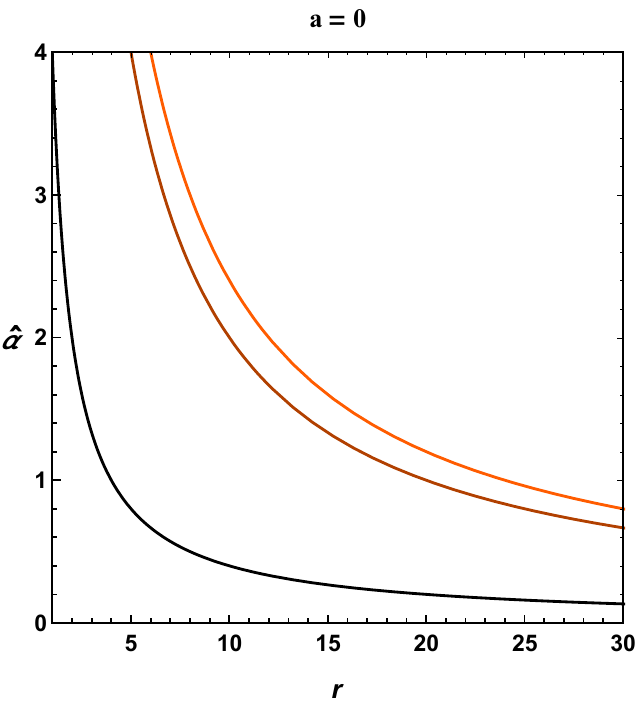} & \includegraphics[scale=0.41]{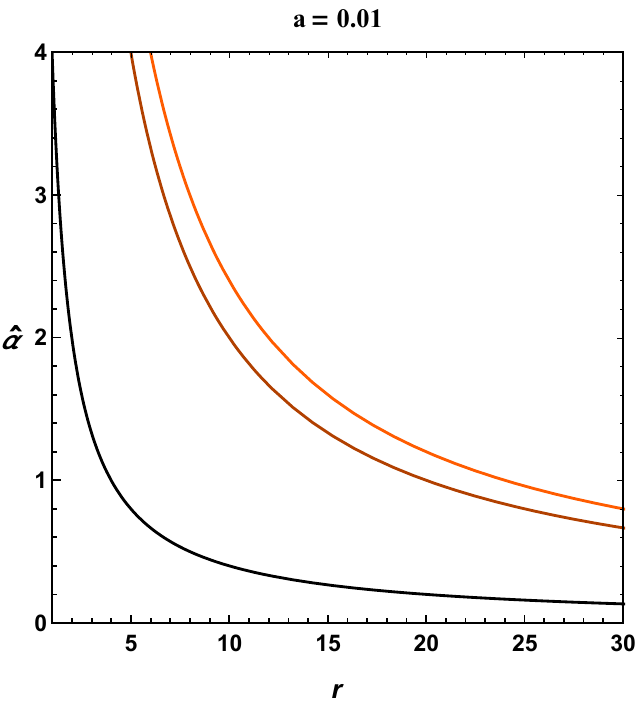} & %
\includegraphics[scale=0.41]{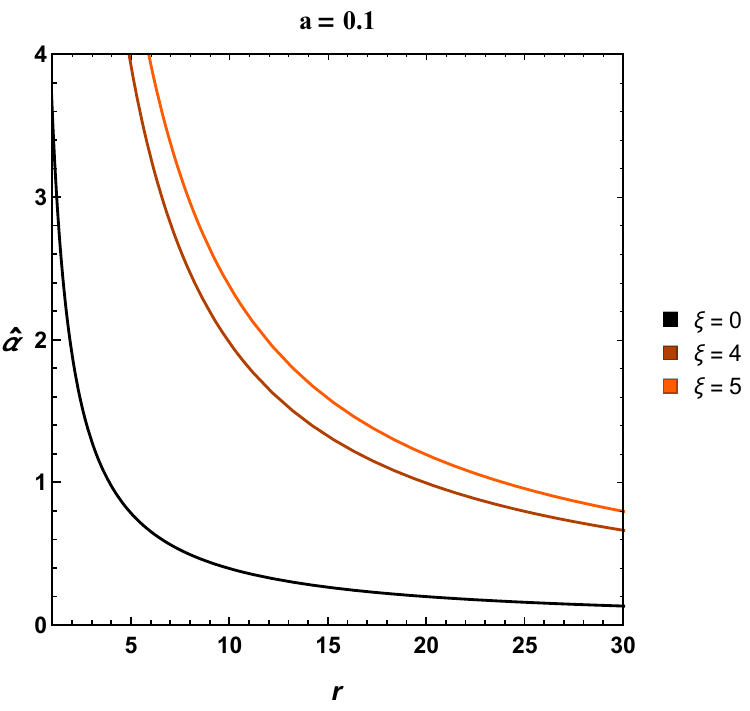}%
\end{tabular}%
\caption{ {\protect\footnotesize Deflection angle for different values of
the acoustic parameter $\protect\xi$ and the spin $a$ and fixed black hole
mass $\left( M=1 \right)$. }}
\label{Defl}
\end{figure}

The metric \eqref{last} describes a slowly rotating acoustic
black hole. We begin by rewriting such a metric in the equatorial plan in
order to get the required the deflection angle of light. Indeed for $\theta
= \frac{\pi}{2} $, we have
\begin{equation}
ds^2=-\left[ F(r)+2a \left(1-F(r) \right)\frac{d\phi}{dt} \right] dt^2+\frac{%
dr^2}{F(r)}+r^2d\phi^2.
\end{equation}
The equations given in \eqref{dphidtau} can be used to
compute the factor $\frac{d\phi}{dt}$ in this equation. The expression for
such a quantity as a function of the impact parameter $b=\frac{L}{E}$, can
be given by
\begin{equation}
\frac{d\phi}{dt}=\frac{F(r)+b a \left(1-F(r) \right)}{br^2-a\left(1-F(r)
\right)}.
\end{equation}
Introducing the two new variables
\begin{align}
dr_{*} &=\frac{dr}{\sqrt{F(r) \left( F(r)+2a\left(1-F(r) \right) \frac{d\phi%
}{dt} \right)}}, \\
f(r_*) &=\frac{r}{\sqrt{F(r)+2a\left(1-F(r) \right) \frac{d\phi}{dt}}},
\end{align}
allows one to derive the optical metric on the equatorial plan for null
geodesics $(ds^2=0)$
\begin{equation}
dt^2=g^{opt}_{mn}dx^m dx^n=dr_*^2+F(r_*)^2d\phi^2.
\end{equation}
We employ the Gauss-Bonnet theorem, which connects the optical geometry and
topology, to compute the deflection angle. Such a theorem is written as
\begin{equation}
\iint_{D_{R}} K_{\mathcal{G}} dS+\oint_{\partial D_{R} }kdt+ \sum n_i=2\pi
\chi \left( D_{R} \right).
\end{equation}
where the non-singular optical region $D_{R}$ has a boundary $\partial D_{R}$%
, the geodesic curvature is $k$, the Gaussian optical curvature is
represented by $K_{\mathcal{G}}$ and
$\chi \left( D_{R}
\right)$ is associated to the Euler characteristic. A geodesic curve,
denoted as $\gamma_R$ can be used to express the geodesic curvature as
\begin{equation}
k \left( \gamma_R \right)= \big\vert \nabla _{\gamma_R} \dot{\gamma_R} %
\big\vert.
\end{equation}
Given that $\gamma_R=R=cte$ is verified by the geodesic $\gamma_R$ \cite%
{def1},
where $R$ is the "geometrical size" of the region
$D_R$, the radial component of $k \left(\gamma_R \right)$ becomes
\begin{equation}
\left( \nabla _{\gamma_R} \dot{\gamma_R} \right)^r= \dot{\gamma_R}^\phi +
\partial_\phi \dot{\gamma_R}^r+ \Gamma^r_{\phi \phi} \left(\dot{\gamma_R}%
^\phi \right)^2.
\end{equation}
Following \cite{def1}, the second term provides
\begin{equation}
\oint_{\partial D_{R} }kdt=\pi+\widehat{\alpha}.
\end{equation}
Furthermore, the jump angles $\alpha_S$ (source) and $\alpha_O$ (observer)
are equal to $\frac{\pi}{2}$ when the geometrical size $R$ of the optical
region $D_R$ approaches infinity. Given that the interior angles are $%
n_S=\pi-\alpha_S$ and $n_O=\pi-\alpha_O$ and by adopting the linear approach
of the light ray, the deflection angle may be defined simply by
\begin{equation}
\widehat{\alpha}=-\int_0^\pi \int_{\frac{b}{\sin \phi}}^\infty K_{\mathcal{G}%
}dS,
\end{equation}
where $dS \simeq rdrd\phi$.
The Gaussian optical curvature
can computed in terms of the Ricci scalar by the following equation
\begin{equation}
K_{\mathcal{G}}=\frac{\mathbf{R}}{2},
\end{equation}
which gives
\begin{equation}
K_{\mathcal{G}} = \frac{18 a b M (\xi +1)}{r^5}-\frac{2 M (\xi +1)}{r^3}.
\end{equation}
The deflection angles is finally expressed as
\begin{equation}
\widehat{\alpha} = \frac{4 M (\xi +1)}{b} -\frac{4 a M (\xi +1)}{b^2} +%
\mathcal{O}\left( M^2, a^2 \right),  \label{DefA}
\end{equation}
where $M$ higher orders are not included. At the limit $\xi \to 0$, the
deflection angle \eqref{DefA} becomes
\begin{equation}
\widehat{\alpha} \simeq \frac{4 M}{b} \mp \frac{4 a M}{b^2},
\end{equation}
where the $\mp$-branches would pertain to the pro-/retrograde motion. The
obtained result matches precisely the LTBH black hole deflection of light
\cite{C2}. To analyse the deflection of light by a slowly rotating acoustic
black hole, we plot the associated behaviors as a function of the impact
parameter $b$ in figure \eqref{Defl} for different values of the acoustic
parameter $\xi$ and the spin $a$.

From such plots, we observe that the acoustic parameter $\xi$ has a strong
impact on the deflection of light. Such behaviour, should indeed be expected
from the shadow study of the previous section. By comparing the different
curves, we notice that the light is strongly deflected for the values $\xi
\geq 4$. Considering the rotation coded in the parameter $a$, only small
deviations occur suggesting that the acoustic parameter might have stronger
influence thant the rotation of the black hole it self.

\section{Conclusions}

\label{VII}

In this paper, we have reviewed the Lense-Thirring acoustic black hole
(LTABH), and probed its near horizon features. We showed that the LTABH
exhibit similarities and differences with respect to its gravitational
analogue, the Lense-Thirring Black Hole (LTBH). For what concerns the metric
function, we showed that the rotation parameter does not affect its roots,
as instead does the acoustic parameter $\xi $, which gives rise to radii
depending on its value. More precisely, we found the existence of an inner
acoustic horizon $r_{ac_{-}}$ and of an outer acoustic horizon $r_{ac_{+}}$
when $\xi >4$. The extremal regime of the LTABH is reached for $\xi =4$,
implying $r_{ac_{-}}=r_{ac_{+}}=4M$. Based on these results, also by means
of various graphics, we have characterized the metric function of the LTABH,
and split its spacetime into four regions.

Furthermore, we have defined an effective potential, and studied its
critical points (in particular, its maxima). We showed that the effective
potential of the LTABH is characterized by two maxima, which are associated
to the photon sphere (of radius $r_{ps}$) and to the acoustic sphere (of
radius $r_{as}$), respectively; such radii have been placed and investigated
carefully, as functions of the acoustic parameter $\xi $ and of the rotation
parameter $a$. We have been therefore able to conclude the LTABH is
characterized by two shadows : the \textit{optical shadow}, denoted as $%
R_{s} $, and the \textit{acoustic shadow} denoted as $R_{as}$. In this
context, the limit $\xi \rightarrow 0$ has provided an important consistency
check of our approach.

By varying the rotation parameter $a$, we noticed that it strongly affects
the radius of the optical shadow, while it has less impact onto the radius
of the acoustic shadow. This fact has also been observed on the shapes of
the shadows themselves : in fact, by varying $a$, we have observed that the
acoustic shadow retains its circular shape, whereas the optical shadow is
shifted rightward. On the other hand, we established that the acoustic
parameter $\xi $ is responsible for the size of the shadows; all in all, one
should expect larger shadows when it comes to an acoustic black hole. All of
such results have also been confirmed by the analyses of the distortions $%
\delta _{s/as}$, as well as of the shadow radi $R_{s/as}$.

By deriving the magnitude of the precession frequency $\Omega $, we also
have investigated the frame dragging effect in the LTABH. Of course, our
results perfectly match the Kerr magnitude in the limit $\xi \rightarrow 0$.
To shed light on the frame dragging, we have set up a parametric description
and varied the relevant parameters accordingly : this allowed us to provide
substantial evidence that $\Omega $ significantly increases near the
acoustic horizons, both in the extremal case and in the non-extremal one;
this is an indication that the frame dragging becomes important near such
regions. Moreover, it has been shown that $\Omega $ becomes null in some
radial regions at the equatorial plan; such regions have analytically been
determined, and we were able to find that in some regions of the parameter
space the orbits of probe particles would not be affected by the frame
dragging. Interestingly, this insensitivity has been traced back to the
presence of the acoustic parameter $\xi $. \textit{Last but not least}, we
have derived and analyzed the deflection of light by the LTABH, conforming
and detailing the results obtained in the shadow analysis.\bigskip

As of today, acoustic black holes have been the object of a vast variety of
studies, both at the theoretical and experimental level. We share the
belief, widely spread within the scientific community, that a deeper
understanding of such black holes may importantly contribute to formulate
models of more and more realistic astrophysical black holes. We can
therefore reasonably state that the rotating class of acoustic black holes,
to which the LTABH introduced in this paper belongs, warrants and deserves
further investigation (for instance, the construction of an analogue-Kerr
black hole still remains an open issue). We leave such intriguing tasks to
future endeavours.

\newpage

\appendix

\section{On the placement of critical radii}

\label{app1} 
\begin{figure}[tbp]
\begin{tabular}{lll}
\includegraphics[scale=0.28]{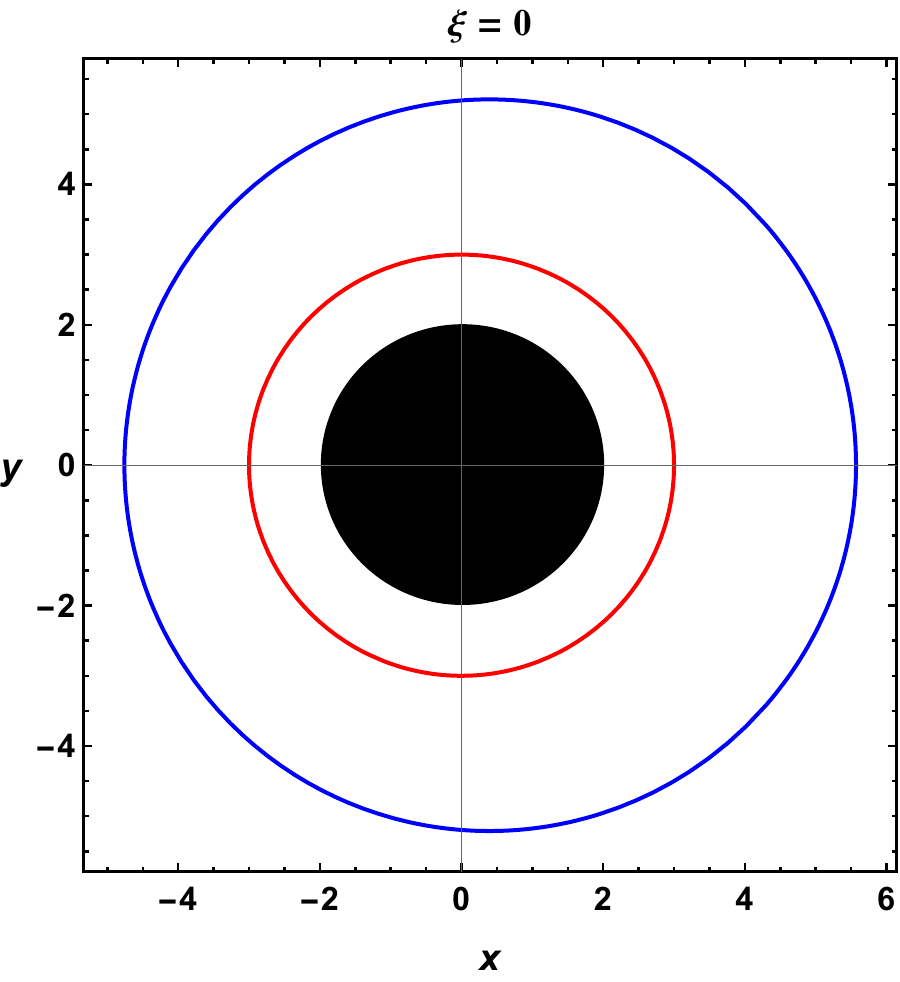} & \includegraphics[scale=0.28]{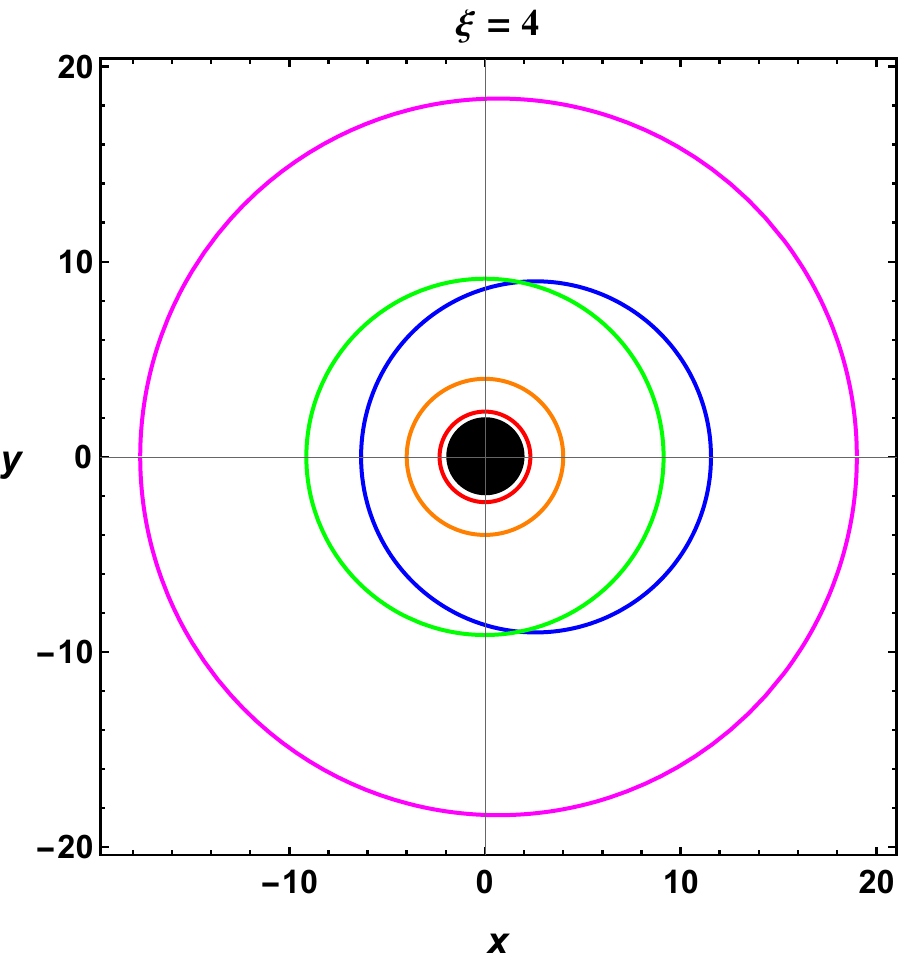} & %
\includegraphics[scale=0.34]{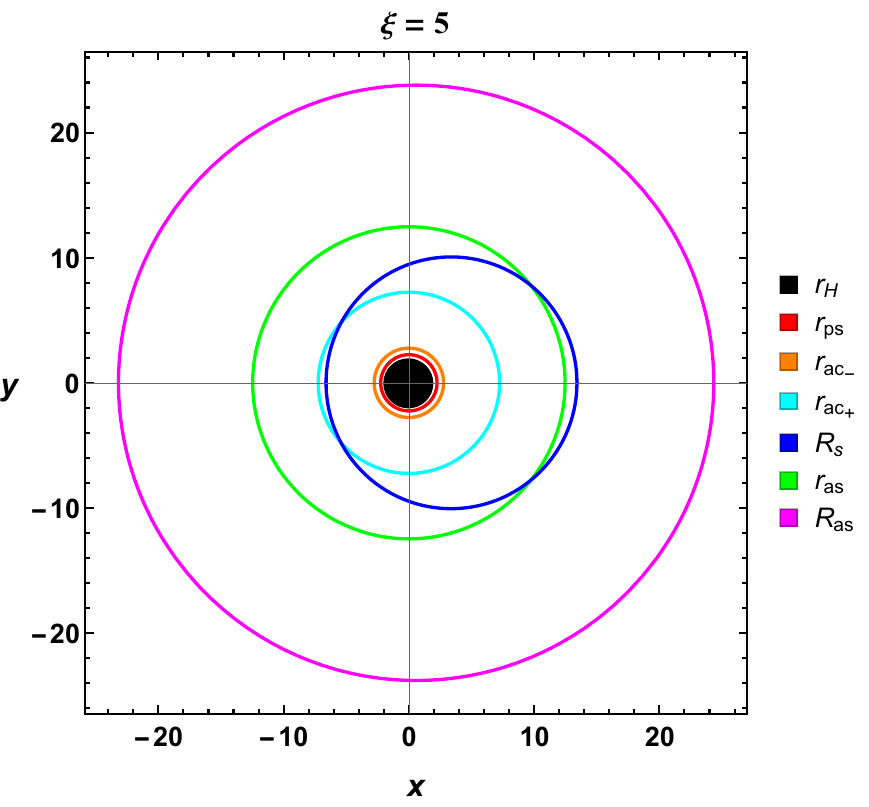}%
\end{tabular}%
\caption{ {\protect\footnotesize Illustration of the critical radii for
different values of of the acoustic parameter $\protect\xi $. Left: Non
acoustic case. Middel: Extremal case of the acoustic black hole. Right:
Acoustic black hole. We take $a=0.2$, $M=1$, $E=%
\mathcal{K}=1$ and $L=20$. }}
\label{ILLL}
\end{figure}

Further investigation could be carried regarding the LTABH who's spacetime is described by the metric of
equation \eqref{last}. Indeed, we have seen that such a spacetime strongly
depend on the value of the acoustic parameter $\xi $. While the horizon radius
is always equal to $2M$, distinctions are also present. For convenience, we
use the following notations
\begin{equation}
r_{1}:=r_{ps},\,\,R_{s}\left( r_{1}\right) =R_{s}\left( r_{ps}\right)
:=R_{s},\quad r_{2}:=r_{as},\,\,R_{s}\left( r_{2}\right) =R_{s}\left(
r_{as}\right) :=R_{as}.
\end{equation}%
In what follows, we discuss and picture each case for a better understanding
of such important spacetime distribution. First, we have the non acoustic
case $\left( \xi =0\right) $ in which the inner and outer acoustic horizon
vanishes. The relevant radii for this particular case are given in table %
\eqref{T1}.
\begin{table}[H]
\caption{{\protect\footnotesize Critical radii values for the non acoustic
case. We take $M=1$, $E=%
\mathcal{K}=1$ and $L=20$.}}
\label{T1}\centering
\begin{tabular}{|c|c|c|c|c|}
\hline
$\xi$ & $r_H$ & $a$ & $r_{ps}$ & $R_s$ \\ \hline
\multirow{3}{*}{$0$} & \multirow{3}{*}{$2$} & $0$ & $3$ & $5.196$ \\
\cline{3-5}
&  & $0.1$ & $3.03$ & $5.248$ \\ \cline{3-5}
&  & $0.2$ & $3.06$ & $5.3$ \\ \hline
\end{tabular}%
\end{table}
Thus, we remark that for such a case the rotation parameter $a$ increases
the photon sphere $r_{ps}$ and the shadow radius $R_{s}$. The extremal
acoustic case in which $\xi =4$ is characterized by a horizon and an
acoustic radius. Both of these quantities does not depend on the rotating
parameter $a$. The other relevant quantities are provided in table \eqref{T2}%
.
\begin{table}[H]
\caption{{\protect\footnotesize Critical radii values for the extremal
acoustic case. We take $M=1$, $E=%
\mathcal{K}=1$ and $L=20$.}}
\label{T2}\centering
\begin{tabular}{|c|c|c|c|c|c|c|c|}
\hline
$\xi$ & $r_H$ & $r_{ac}$ & $a$ & $r_{ps}$ & $r_{as}$ & $R_s$ & $R_{as}$ \\
\hline
\multirow{3}{*}{$4$} & \multirow{3}{*}{$2$} & \multirow{3}{*}{$4$} & $0$ & $%
2.299$ & $8.7$ & $8.62$ & $18.35$ \\ \cline{4-8}
&  &  & $0.1$ & $2.31$ & $8.92$ & $9.24$ & $18.67$ \\ \cline{4-8}
&  &  & $0.2$ & $2.32$ & $9.13$ & $10$ & $18.98$ \\ \hline
\end{tabular}%
\end{table}
It is clear from such a table that all the critical radii values increases
with the rotating parameter $a$. By comparing these values with the case
where $\xi =0$, it is clear that the impact of $\xi $ might be different on
each quantity. For instance, we remark that the photon sphere $r_{ps}$
decreased while the shadow radius $R_{s}$ increased. In the case where $\xi
\geq 4$, there is an inner and outer acoustic horizon which are $r_{ac_{-}}$
and $r_{ac_{+}}$ respectively. In table \eqref{T3}, we specify the relevant
radii for different values of the rotating parameter $a$.
\begin{table}[H]
\caption{{\protect\footnotesize Critical radii values for the acoustic case.
We take $M=1$, $E=%
\mathcal{K}=1$ and $L=20$.}}
\label{T3}\centering
\begin{tabular}{|c|c|c|c|c|c|c|c|c|}
\hline
$\xi$ & $r_H$ & $r_{ac_{-}}$ & $r_{ac_{+}}$ & $a$ & $r_{ps}$ & $r_{as}$ & $%
R_s$ & $R_{as}$ \\ \hline
\multirow{3}{*}{$5$} & \multirow{3}{*}{$2$} & \multirow{3}{*}{$2.77$} & %
\multirow{3}{*}{$7.24$} & $0$ & $2.23$ & $12.05$ & $9.47$ & $23.78$ \\
\cline{5-9}
&  &  &  & $0.1$ & $2.24$ & $12.26$ & $10.39$ & $24.12$ \\ \cline{5-9}
&  &  &  & $0.2$ & $2.25$ & $12.47$ & $11.63$ & $24.47$ \\ \hline
\end{tabular}%
\end{table}
From such a table, we can confirm that the photon sphere $r_{ps}$ decreases
with the increase the parameter $\xi $ while all the other quantities
increases. The impact of the rotating parameter $a$ is the same for all
quantities as it can be observed from the tables. It can be remarked that the photon sphere $r_{ps}$ has decreased more in size compared to the extremal acoustic case while the other quantities have increased. In figure \ref{ILLL}, we report each of the critical radii for all the cases discussed above.
It is worth noting that the interaction Photon-Phonon is not taken into
account in this paper; we leave the investigation of such an interaction for
further future work.

\section{The matrix $\mathcal{G}_{\protect\mu \protect\nu }$}

\label{app 2} The explicit $\mathcal{G}_{\mu \nu }$ matrix is
\begin{equation}
\mathcal{G}_{\mu \nu }=%
\begin{pmatrix}
\mathcal{G}_{tt} & \mathcal{G}_{tr} & \mathcal{G}_{t\theta } & \mathcal{G}%
_{t\phi } \\
\mathcal{G}_{rt} & \mathcal{G}_{rr} & \mathcal{G}_{r\theta } & \mathcal{G}%
_{r\phi } \\
\mathcal{G}_{\theta t} & \mathcal{G}_{\theta r} & \mathcal{G}_{\theta \theta
} & \mathcal{G}_{\theta \phi } \\
\mathcal{G}_{\phi t} & \mathcal{G}_{\phi r} & \mathcal{G}_{\phi \theta } &
\mathcal{G}_{\phi \phi }%
\end{pmatrix}%
,
\end{equation}%
with
\begin{align}
\mathcal{G}_{tt}& =\frac{c_{s}}{\sqrt{c_{s}^{2}-v_{\alpha }v^{\alpha }}}%
\,g_{tt}\left( c_{s}^{2}-v_{i}v^{i}\right) , \\
\mathcal{G}_{tr}& =\frac{c_{s}}{\sqrt{c_{s}^{2}-v_{\alpha }v^{\alpha }}}%
\,\left( -v_{t}v_{r}+g_{tr}(c_{s}^{2}-v_{i}v^{i})\right) , \\
\mathcal{G}_{t\theta }& =\frac{c_{s}}{\sqrt{c_{s}^{2}-v_{\alpha }v^{\alpha }}%
}\,\left( -v_{t}v_{\theta }+g_{t\theta }(c_{s}^{2}-v_{i}v^{i})\right) , \\
\mathcal{G}_{t\phi }& =\frac{c_{s}}{\sqrt{c_{s}^{2}-v_{\alpha }v^{\alpha }}}%
\,\left( -v_{t}v_{\phi }+g_{t\phi }(c_{s}^{2}-v_{i}v^{i})\right) , \\
\mathcal{G}_{rt}& =\frac{c_{s}}{\sqrt{c_{s}^{2}-v_{\alpha }v^{\alpha }}}%
\,\left( -v_{r}v_{t}+g_{rt}(c_{s}^{2}-v_{t}v^{t})\right) , \\
\mathcal{G}_{rr}& =\frac{c_{s}}{\sqrt{c_{s}^{2}-v_{\alpha }v^{\alpha }}}%
\,\left( g_{rr}(c_{s}^{2}-v_{\alpha }v^{\alpha })+v_{r}v_{r}\right) ,
\end{align}
\begin{align}
\mathcal{G}_{r\theta }& =\frac{c_{s}}{\sqrt{c_{s}^{2}-v_{\alpha }v^{\alpha }}%
}\,\left( g_{r\theta }(c_{s}^{2}-v_{\alpha }v^{\alpha })+v_{r}v_{\theta
}\right) , \\
\mathcal{G}_{r\phi }& =\frac{c_{s}}{\sqrt{c_{s}^{2}-v_{\alpha }v^{\alpha }}}%
\,\left( g_{r\phi }(c_{s}^{2}-v_{\alpha }v^{\alpha })+v_{r}v_{\phi }\right) ,
\\
\mathcal{G}_{\theta t}& =\frac{c_{s}}{\sqrt{c_{s}^{2}-v_{\alpha }v^{\alpha }}%
}\,\left( -v_{\theta }v_{t}+g_{\theta t}(c_{s}^{2}-v_{t}v^{t})\right) , \\
\mathcal{G}_{\theta r}& =\frac{c_{s}}{\sqrt{c_{s}^{2}-v_{\alpha }v^{\alpha }}%
}\,\left( g_{\theta r}(c_{s}^{2}-v_{\alpha }v^{\alpha })+v_{\theta
}v_{r}\right) , \\
\mathcal{G}_{\theta \theta }& =\frac{c_{s}}{\sqrt{c_{s}^{2}-v_{\alpha
}v^{\alpha }}}\,\left( g_{\theta \theta }(c_{s}^{2}-v_{\alpha }v^{\alpha
})+v_{\theta }v_{\theta }\right) , \\
\mathcal{G}_{\theta \phi }& =\frac{c_{s}}{\sqrt{c_{s}^{2}-v_{\alpha
}v^{\alpha }}}\,\left( g_{\theta \phi }(c_{s}^{2}-v_{\alpha }v^{\alpha
})+v_{\theta }v_{\phi }\right) , \\
\mathcal{G}_{\phi t}& =\frac{c_{s}}{\sqrt{c_{s}^{2}-v_{\alpha }v^{\alpha }}}%
\,\left( -v_{\phi }v_{t}+g_{\phi t}(c_{s}^{2}-v_{t}v^{t})\right) , \\
\mathcal{G}_{\phi r}& =\frac{c_{s}}{\sqrt{c_{s}^{2}-v_{\alpha }v^{\alpha }}}%
\,\left( g_{\phi r}(c_{s}^{2}-v_{\alpha }v^{\alpha })+v_{\phi }v_{r}\right) ,
\\
\mathcal{G}_{\phi \theta }& =\frac{c_{s}}{\sqrt{c_{s}^{2}-v_{\alpha
}v^{\alpha }}}\,\left( g_{\phi \theta }(c_{s}^{2}-v_{\alpha }v^{\alpha
})+v_{\phi }v_{\theta }\right) , \\
\mathcal{G}_{\phi \phi }& =\frac{c_{s}}{\sqrt{c_{s}^{2}-v_{\alpha }v^{\alpha
}}}\,\left( g_{\phi \phi }(c_{s}^{2}-v_{\alpha }v^{\alpha })+v_{\phi
}v_{\phi }\right) .
\end{align}

\end{document}